\documentclass{aa}
\usepackage{graphicx}
\usepackage{latexsym}

\newcommand{\km}{\,\mbox{km}\,\mbox{s}^{-1}}
\newcommand{\bsao}{Bull. Spec. Astrophys. Obs.}
\def\Ha{\hbox{H$_\alpha$\,}}
\def\Hb{\hbox{H$_\beta$\,}}
\def\Rc{$\mbox{R}_c$\,}
\def\K{$\mbox{K}'$\,}
\def\pak{$PA_{dyn}$\,}

\begin{document}

\setlength{\tabcolsep}{0.1cm}

   \title{Structure and kinematics of candidate double-barred galaxies.\thanks{
   Based on observations carried out at the 6m telescope
of the Special Astrophysical Observatory of the Russian Academy of Sciences,
operated under the financial support of the Science Department of Russia
(registration number 01-43), at the 2.1m telescope of the {\it Observatorio
Astron\'onico Nacional}, San Pedro Martir, M\'exico, and  from the data
archive of the NASA/ESA Hubble Space Telescope at the Space Telescope Science
Institute. STScI is operated by the association of Universities for Research
in Astronomy, Inc. under NASA contract NAS  5-26555.}
}

   \author{A.V. Moiseev\inst{1}\thanks{Guest User, Canadian Astronomy Data
   Center, which is operated by the Herzberg Institute of Astrophysics, National
                       Research Council of Canada.}
    \and J.R. Vald\'es\inst{2} \and  V.H. Chavushyan\inst{2}
          }

   \offprints{A.V. Moiseev \email{moisav@sao.ru}}

   \institute{Special Astrophysical Observatory, Nizhnij Arkhyz,
 Karachaevo-Cherkesia, 369167, Russia
         \and
     Instituto Nacional de Astrof\'{\i}sica Optica y Electr\'onica,
     Apartado Postal 51 y 216, C.P. 72000, Puebla, Pue., M\'exico
             }

   \date{Received / accepted }

   \abstract{Results of optical and NIR spectral and photometric observations
of a sample of candidate double-barred galaxies are presented.
Velocity fields and velocity dispersion maps of stars and ionized
gas, continuum and emission-line images were constructed from
integral-field spectroscopy observations carried out at the 6m
telescope (BTA) of SAO RAS, with the MPFS spectrograph and the
scanning Fabry-Perot Interferometer. \object{NGC 2681} was also observed
with the long-slit spectrograph of the  BTA. Optical and NIR
images were obtained at the BTA and at the 2.1m telescope (OAN,
M\'exico). High-resolution images were retrieved from the  HST
data archive. Morphological and kinematic features of all 13
sample objects  are described in detail. Attention is focused on
the interpretation of observed non-circular motions of gas and
stars in circumnuclear (one kiloparsec-scale) regions. We have
shown first of all that these motions are caused by the
gravitational potential of a large-scale bar. \object{NGC 3368} and
\object{NGC 3786} have nuclear bars only, their isophotal twist at larger
radii being connected with the bright spiral arms. Three cases of
inner polar disks in our sample (\object{NGC 2681}, \object{NGC 3368} and
\object{NGC 5850}) are considered. We found ionized-gas counter-rotation
in the central kiloparsec of the lenticular galaxy \object{NGC 3945}.
Seven galaxies (\object{NGC 470}, \object{NGC 2273}, \object{NGC 2681}, \object{NGC 3945}, \object{NGC 5566},
\object{NGC 5905}, and \object{NGC 6951}) have inner mini-disks nested in
large-scale bars. Minispiral structures occur often in these
nuclear disks. It is interesting that the majority of the
observed, morphological and kinematical, features in the sample
galaxies can be explained without the secondary bar hypothesis.
Thus we suggest that a dynamically independent secondary bar is a
rarer  phenomenon than  follows from isophotal analysis of the
images only.

             }

\titlerunning{Structure and kinematics of candidate  double-barred
galaxies.}
\authorrunning{Moiseev et al.}
\maketitle

\section{Introduction}

\label{intro}

More than a quarter of a century ago de Vaucouleurs (\cite{vauc})
found that an inner bar-like structure was nested in the
large-scale bar in the optical image of \object{NGC 1291}. The first
systematic observational study of this phenomenon was made by Buta
\& Crocker (\cite{butacr}), who published a list of 13 galaxies,
where the secondary (inner) bar was arbitrarily oriented with
respect to the primary (outer) one. In the following years the
isophotal analysis technique made it possible  to detect secondary
bars in optical (Wozniak et al., \cite{woz}; Erwin \& Sparke,
\cite{erw99}; Erwin \& Sparke, \cite{erw03}) and NIR images of
barred galaxies (Friedli et al., \cite{fri96}; Jungwiert et al.,
\cite{jung}; Laine et al., \cite{laine02}). Up to now, we have
found 71 candidate  double-barred galaxies in the literature (see
references in Moiseev, \cite{mois01b}). A new revised catalog of
67 barred galaxies including 50 double-barred ones is presented by
Erwin (\cite{erwcat}). Although various theoretical studies exist,
the secondary bar dynamics is still far from being well
understood. Shlosman et al. (\cite{shlosman89}) have shown that an
additional circumnuclear bar may be formed as a result of
instabilities of the gaseous ring in the Inner Lindblad Resonance
(ILR) within a large-scale bar. Heller et al. (\cite{heller01})
present a detailed simulation of this process, which,
unfortunately, results in the appearance of an elliptical gaseous
ring instead of the ``real'' stellar-gaseous bar. Maciejewski \&
Sparke (\cite{mac00}) found some families of closed orbital loops,
able to support both bars. Similar independently rotating
structures sometimes appear in stellar-gaseous simulations of
galactic disks (Pfenninger \& Norman, \cite{pfenninger}; Friedli
\& Martinet, \cite{frimar}). New hydrodynamical simulations of the
gas behavior in double bars, with an analysis of the problems
arising during modeling and interpretation of the double-barred
structures are presented by Maciejewski et al. (\cite{mac02}) and
Shlosman \& Heller (\cite{shlosman02}).

Various observational data indicate that we are probably seeing
 a new structural feature of barred galaxies in the
case of double bars. However, the majority of observational
results has been based on photometric measurements only, when an
elongated elliptical structure appears inside the large-scale bar.
Moreover, using the isophotal analysis formalism (Wozniak et al.,
\cite{woz}) has allowed  even ``triple bars'' (Friedli et al.,
\cite{fri96}; Jungwiert et al., \cite{jung}; Erwin \& Sparke,
\cite{erw99}) without any arguments about the dynamical behavior
of such complex stellar systems. At the same time, the structures
seen in direct images can be explained in a less exotic manner,
without using the hypothesis of double or triple bars. Many
structural features would be able to create an illusion of a
secondary bar (Friedli et al., \cite{fri96}; Moiseev,
\cite{mois01b}; Petitpas \& Wilson, \cite{petit}), such as for
example, an oblate bulge, a complex distributions of dust and star
formation regions in the circumnuclear region, or elliptical rings
at the ILRs of a primary bar (Shaw et al., \cite{shaw}). Therefore
new observational data are required  to test our understanding of
the secondary bar phenomenon.

Kinematic data (the stellar and gas  line-of-sight velocities and
velocity dispersion) over the circumnuclear  regions of barred
galaxies are of crucial importance for this kind of studies.
Recently, Emsellem et al. (\cite{emsellem01}) have studied the
kinematics of double-barred galaxy candidates and reach a
conclusion on dynamical decoupling of the circumnuclear regions in
\object{NGC 1097}, \object{NGC  1808} and \object{NGC 5728}.
This conclusion is based on the
fact that peaks of the relative line-of-sight velocities and
stellar velocity dispersion drops  are seen in the circumnuclear
regions from long-slit cross-sections. Wozniak (\cite{woz99}) has
supposed that ``counter-rotation'' of stars, observed in the
nuclear region of \object{NGC 5728}, may be associated with secondary bar
influence on stellar kinematics. All these features observed in
the one-dimensional data may be explained without using the
hypothesis of secondary bars. This  may be related to the specific
mass distributions in the inner kiloparsec and to non-circular
motions of stars and gas within the primary bar only
($x_2$-families of stellar orbits lead to  ``counter-rotation''
properties). Also Wozniak et al. (\cite{woz03}) in their
self-consistent N-body simulations have shown that the velocity
dispersion drops observed by Emsellem et al. (\cite{emsellem01})
can be independent on the presence of a secondary bar and are
reproduced in the model of a single-barred galaxy.

Since the motions of stars and gas inside the bar region are
strongly non-circular, and such objects are non-axisymmetric by
definition,  panoramic (also named 2D, 3D or integral-field)
spectroscopy should give us  more detailed information on
circumnuclear kinematics in compared to  ``classical'' (long-slit)
spectroscopy. Integral-field spectroscopy makes it possible to map
two-dimensional fields of line-of-sight velocities and velocity
dispersion. Therefore, during 2000-2002 we have carried out a
series of observational programs on 2D-kinematics and morphology
of double-barred galaxies candidates. The main goal was to find an
answer to the question -- \textit{Are the secondary bars
dynamically decoupled systems?}

This problem is complicated and ambiguous because of the lack of
theoretical studies, where the ``observed'' velocity fields in the
double-barred galaxies had been modeled. It is important to
collect a full sample of candidate double-barred galaxies with
observable 2D-kinematics of stars and gas. These observations must
be used to search for common dynamical (kinematic) features in
these kinds of galaxies. Recently, Shlosman, \& Heller
(\cite{shlosman02}) have written: ``..clearly, the most promising
method in detecting the nuclear bars is two-dimensional
spectroscopy of the central kiloparsec, which can reveal the
underlying kinematics''.

Our observational program was finished in 2002. Independently
similar observations  were started at the 3.6m CFHT  using the
integral-field spectrographs OASIS and TIGER. Recently, the
velocity fields of stars and ionized gas in some double-barred
candidates have been presented by Emsellem \& Friedli
(\cite{eric00}), Emsellem et al. (\cite{emsellem01b}) and Emsellem
(\cite{eric470}). The velocity fields of the molecular gas were
also investigated in double-barred candidates \object{NGC 2273} and
\object{NGC 5728} (Petitpas \& Wilson, \cite{petit}) and \object{NGC 4303}
(Schinnerer et al., \cite{eva}).

In this work, we present the observational data (integral-field
spectroscopy and NIR/optical surface photometry) for 13 galaxies
and discuss the problem of central region dynamical decoupling.
The paper is structured as follows. In Sect.~\ref{sec_obs} we
describe the observations and data reduction technique. The
methods of analysis of the velocity fields and surface brightness
distributions are considered in Sect.~\ref{method}. In
Sect.~\ref{atlas0} we present the notes for individual observed
galaxies and discuss their common properties in Sect.~\ref{disc};
a short  conclusion is drawn in Sect.~\ref{concl}.

\section{Observations and data reduction}

\label{sec_obs}

We have compiled our sample from Moiseev (\cite{mois01b}) by
applying the following restrictions: $\delta>0$, and the size of
the major axis of the probable secondary bar is smaller than or
equal to the MPFS field of view ($16''\times 15''$). We have
obtained full observational data for 13 galaxies, that is for
about one third of the candidates in the Northern sky.

\subsection{Integral-field spectroscopy with the MPFS}

Circumnuclear regions of all the galaxies were observed at the 6m
telescope of the Special Astrophysical Observatory, Russian
Academy of Sciences (SAO RAS) with the Multipupil Integral Fiber
Spectrograph (MPFS, Afanasiev et al., \cite{mpfs}). A description
of the spectrograph is also  available through the Internet at
SAO WEB-page \verb*"http://www.sao.ru/hq/lsfvo/devices.html". The
spectrograph takes simultaneous spectra from 240 spatial elements
(constructed in the form of square lenses) that form an array of
$16\times 15$ elements on the sky. The field of view is
$16''\times 15''$, and the angular size is $1''$/element.

Three galaxies, \object{NGC 2273}, \object{NGC 3368}, and \object{NGC 3945} were observed in
two different positions of the multilens array on the sky, with
each frame containing a galactic nucleus. After preliminary data
reduction, the ``data cubes'' were combined. The resulting field
of view was $16''\times18''$ for \object{NGC 2273}, $18''\times18''$ for
\object{NGC 3368} and $23''\times23''$ for \object{NGC 3945}. Together with target
spectra, we took a night-sky spectrum from an area located
4\farcm5 from the center of the field of view. The detector was a
TK1024 $1024\times1024$ pixels CCD array. The spectrograph
reciprocal dispersion was $1.35$\AA/pixel and the spectral
resolution was about $4$\AA. The log of MPFS observations is given
in Table~\ref{tab_mpfs}. It contains the names of galaxies, dates
of the observations, the total exposures $T_{exp}$, the spectral
range $\Delta\lambda$ and the seeing quality during the exposures.

We reduced the observations by using the software developed at the
SAO RAS by V.L. Afanasiev and running in the IDL environment. The
primary reduction included bias subtraction, flat-fielding,
cosmic-ray hits removal, extraction of individual spectra from
the CCD frames, and their wavelength calibration using a
spectrum of a He-Ne-Ar lamp. Subsequently, we subtracted the
night-sky spectrum from the galaxy spectra. The spectra of
spectrophotometric standard stars were used to convert the fluxes
into absolute energies.

The reduced spectra were represented as ``data cubes'': each
spatial element of the two-dimensional field has an individual
1024-channel spectrum. We constructed the maps of line intensity
and light-of-sight velocity fields in the \Hb, [O~III]
$\lambda4959,5007$~{\AA} and/or [N~II]$\lambda6548,6583$~{\AA}
lines by means of the Gaussian fitting of the emission line
profiles. A double-gaussian model was used for doublet separation.
The absolute accuracy of the velocity determination, evaluated
from the air-glow lines wavelengths, was about $10-15\km$. For the
continuum map construction we summed the fluxes in the spectral
ranges free of  emission lines ($5600-5800$~\AA\, for ``green''
and $6350-6450$\AA\, for ``red'' spectra).

The line-of-sight velocity and velocity dispersion fields for the
stellar component were constructed by means of a cross-correlation
technique, modified for the MPFS data (Moiseev, \cite{mois01a}).
The used spectral region included some stellar absorption
features; MgI$\lambda5175$~\AA,\, FeI$\lambda5229$~\AA,
FeI+CaI$\lambda5270$~\AA,\, and NaI$\lambda5893$~\AA\ among
others. The spectra of G8-K3 giant stars and the twilight sky,
observed during the same nights, were obtained as velocity
templates. To take into account the variations of the instrumental
contour (see Moiseev, \cite{mois01a})the observations of the
template, the images of stars were out-of-focus to fill the MPFS
field of view. The accuracies were of $\sim10\km$, and $10-20\km$
respectively for the velocity and velocity dispersion
determinations  (the errors are dependent on the cross-correlation
peak amplitudes).

\subsection{Observations with the Fabry-Perot Interferometer}

\label{sec22}

Six galaxies with bright emission lines were observed at the 6m
telescope with the scanning Fabry-Perot Interferometer (IFP). The
Queensgate interferometer ET-50 was used in 235th interference
order (for the \Ha line). The free spectral range between
neighboring orders (interfringe) was about 28\,\AA\, ($1270\km$).
For preliminary monochromatization, a set of narrow-band filters,
with a $FWHM=12-18$~\AA\, and centered on the spectral region
containing redshifted \Ha or [N~II]$\lambda6583$~\AA\, emission
lines were used. During the observations we successively take 32
interferometric images of an object with different gaps between
IFP plates. The spectral channels have a width of $\delta
\lambda\approx0.9 $\AA\, ($\sim40\km$), the spectral resolution
(the width of instrumental contours) was $FWHM\approx 2.5$~\AA\,
$(\sim110\km)$. The detector was a CCD TK1024 ($1024\times1024$
pixels). The CCD was operated with $2\times2$ binning for
reading-out time economy. Therefore each spectral channel has a
$512\times512$ pixel format.

In February and March, 2000, the IFP was installed into the
parallel beam, inside the focal reducer (Dodonov et al.,
\cite{dodo}; Moiseev, \cite{mois00}). The resulting focal ratio at
the prime focus of 6m telescope was $F/2.4$, with a field of view
of 5\farcm8 and a scale of $0.68''/\mbox{px}$.

In September and November 2000, the observations were carried out
with the new multi-mode focal reducer SCORPIO. A brief description
of this device is given on Internet
(\verb*"http://www.sao.ru/hq/moisav/scorpio/scorpio.html"), also
the IFP mode in SCORPIO is described by   Moiseev (\cite{mois02}).
With this configuration, the focal ratio was $F/2.9$, with a field
of view of 4\farcm8 and a scale of $0.56''/\mbox{px}$. The log of
the IFP observations is presented in Table~\ref{tab_ifp}.

To reduce the interferometric observations we used a custom
development software (Moiseev, \cite{mois02}), running in the IDL
environment. After primary reduction (bias, flat-fielding etc.),
 removing the night-sky emission lines and converting to the wavelength scale, the data
were presented as ``data cubes''. In such a cube, to every pixel
of the $512\times512$ field, a 32-channel spectrum is attached.
The ``data cubes'' were smoothed by Gaussians with a width, in the
spectral domain, of $FWHM=1.5$ channel, and in the spatial
(sky-plane) domain of $FWHM=2-3$ pixels under the ADHOC
package\footnote{ ADHOC software was  written by J. Boulesteix
(Observatoire de Marseille). See
\texttt{http://www-obs.cnrs-mrs.fr/ADHOC/adhoc.html})}. The
velocity fields of the ionized gas, and the images in the emission
line were mapped by means of a Gaussian fitting of the emission
line profiles. Moreover, we created the images of the galaxies in
the ``red'' continuum near to the emission lines.

\begin{table}
\caption{Log of MPFS observations.}
\label{tab_mpfs}
\begin{center}
\begin{tabular}{rrrcr}
\hline
\hline
Name & Date & $T_{exp}$, sec & $\Delta\lambda$,\AA& Seeing \\\hline
\object{NGC 470}  & 01/09/2000 & $4\times1200$ & $4820-6190$ & 2.8$''$\\
\object{NGC 2273} & 01/12/2000 & $6\times1200$ & $4770-6140$ & 2.5$''$\\
\object{NGC 2681} & 30/03/2000 & $4\times1200$ & $4840-6210$ & 2.0$''$\\
                  & 09/03/2002 & $3\times900$ & $5800-7170$ & 1.7$''$\\
\object{NGC 2950} & 27/03/2000 & $3\times1200$ & $4840-6210$ & 2.0$''$\\
\object{NGC 3368} & 27/03/2000 & $7\times1200$ & $4840-6210$ & 2.5$''$\\
\object{NGC 3786} & 27/03/2000 & $4\times1200$ & $4840-6210$ & 2.5$''$\\
\object{NGC 3945} & 01/12/2000 & $6\times1200$ & $4770-6140$ & 2.5$''$\\
                  & 13/05/2002 & $2\times1200$ & $4700-6070$ & 1.5$''$\\
                  & 13/05/2002 & $3\times1200$ & $5900-7270$ & 1.5$''$\\
\object{NGC 4736} & 11/05/2000 & $3\times1200$ & $4840-6210$ & 1.8$''$\\
\object{NGC 5566} & 10/05/2000 & $3\times1200$ & $4840-6210$ & 2.3$''$\\
\object{NGC 5850} & 13/08/2001 & $3\times1200$ & $5710-7080$ & 2.7$''$\\
                  & 13/05/2002 & $3\times1200$ & $4700-6070$ & 1.2$''$\\
\object{NGC 5905} & 30/03/2000 & $8\times1200$ & $4840-6210$ & 2.0$''$\\
\object{NGC 6951} & 03/09/2000 & $5\times1200$ & $4840-6210$ & 2.2$''$\\
\object{NGC 7743} & 02/09/2000 & $3\times1200$ & $4840-6210$ & 2.0$''$\\\hline
\end{tabular}
\end{center}
\end{table}

\begin{table}
\caption{Log of IFP observations.}
\label{tab_ifp}
\begin{center}
\begin{tabular}{rrrrcr}\hline\hline
Name & Date & $T_{exp}$, sec & Line & Seeing \\\hline
\object{NGC 470}  & 02/11/2000 & $32\times200$ & \Ha & 2.1$''$\\
\object{NGC 2273} & 03/11/2000 & $32\times200$ & \Ha & 2.0$''$\\
\object{NGC 3368} & 28/02/2000 & $32\times150$ & \Ha & 2.7$''$\\
                  & 28/02/2000 & $32\times200$ & [N~II] & 3.5$''$\\
\object{NGC 3945} & 03/11/2000 & $32\times190$ & [N~II] & 2.0$''$\\
\object{NGC 4736} & 02/03/2000 & $32\times150$ & \Ha & 2.7$''$\\
                  & 02/03/2000 & $32\times180$ & [N~II] & 3.5$''$\\
\object{NGC 6951} & 23/09/2000 & $32\times120$ & \Ha & 2.5\\
                  & 23/09/2000 & $32\times120$ & [N~II] & 2.5$''$\\
\hline
\end{tabular}
\end{center}
\end{table}

\subsection{NIR photometry at 2.1m telescope}

The NIR surface photometry of the sample galaxies was carried out
with the CAMILA camera (Cruz-Gonzalez et al., \cite{camila}) in
the JH\K bands, on the 2.1m telescope of the {\it Observatorio
Astron\'omico Nacional}, San Pedro Martir, M\'exico. CAMILA is
equipped with a NICMOS3 $256\times256$ pixels detector.
Observations were carried out with two telescope focuses, with
different focal ratios. In the $F/13.5$ mode the field of view was
1\farcm3 and the pixel size 0\farcs30. The $F/4.5$ mode provided a
field of view of 3\farcm6 with a scale of $0\farcs85/\mbox{px}$.
The log of the NIR observations is given in Table~\ref{tab_nir}.

Each observation consisted of a sequence of object and sky exposures,
with the integration time of an individual exposure limited by the
background level, which was kept well below the non-linear regime of
the control electronics.
The nearby sky frames were taken with the same exposure times and at
adjacent positions to the object, through an observing sequence that
alternates the object and sky frames.

The routines from Image Reduction and Analysis Facility
(\mbox{IRAF}) were used in the reduction and analysis of all the
data. The image processing involved subtraction of sky frames,
division by flat field frames, registration of the images to a
common coordinate system and stacking all the images in a filter.
Bias subtraction was carried out during the data acquisition. Some
of the ima\-ges had a horizontal band at the joints of the
individual quadrants of the \mbox{NICMOS 3} array detector. We
eliminated this feature by subtracting a 255$\times$1 pixels
median smoothed image from the original images.

\begin{table}
\caption{Log of CAMILA observations.} \label{tab_nir}
\begin{center}
\begin{tabular}{rrllr}\hline\hline
Name & Date & Mode & Filters & Seeing \\
\hline
\object{NGC 2273} & 09/03/2001 & $(F/4.5)$& J, H, \K & $3.5''$\\
\object{NGC 2681} & 09/03/2001 & $(F/4.5)$& J, H, \K & $2.8''$\\
                  & 18/03/2000 & $(F/13.5)$& J, H, \K & $1.5''$\\
\object{NGC 2950} & 12/03/2001 & $(F/4.5)$& J, H & $3.0''$\\
                  & 18/03/2000 & $(F/13.5)$& J, H, \K & $1.5''$\\
\object{NGC 3368} & 12/03/2001 & $(F/4.5)$& J, H & $2.2''$\\
                  & 18/03/2000 & $(F/13.5)$& J, H, \K & $1.5''$\\
\object{NGC 3786} & 19/03/2000 & $(F/13.5)$& J, H & $1.5''$\\
\object{NGC 3945} & 09/03/2001 & $(F/4.5)$& J, H & $4.0''$\\
                  & 18/03/2000 & $(F/13.5)$& J, H, \K & $1.5''$\\
\object{NGC 4736} & 20/03/2000 & $(F/4.5)$& J, H, \K & $1.5''$\\
\object{NGC 5566} & 18/03/2000 & $(F/13.5)$& J, H, \K & $1.5''$\\
                  & 12/03/2001 & $(F/4.5)$& J, H & $1.6''$\\
\object{NGC 5850} & 18/03/2000 & $(F/13.5)$& J, H, \K & $1.5''$\\
\object{NGC 5905} & 19/03/2000 & $(F/13.5)$&J, H & $2.5''$\\
\hline
\end{tabular}
\end{center}
\end{table}

\subsection{Direct imaging on the 6m telescope}

Optical direct imaging observations of some galaxies were carried
out with the focal reducer SCORPIO (Sect.~\ref{sec22}) at the 6m
telescope. The detector (CCD TK1024, $1024\times1024$ pixels)
provided a field of view of 4\farcm8 with a pixel scale of
0\farcs28. The log of the observations is presented in
Table~\ref{tab_bta}. We used 3 filters: V (Johnson), \Rc (Cousins)
and a medium-band red filter (SED755) for the continuum imaging of
the objects (without  any pollution from the emission lines of the
object). The SED755 filter has a width of $FWHM=220$~\AA\, and is
centered at $7560$~\AA. The data reduction included bias and
flat-field corrections, and cosmic hits removing.

\subsection{Hubble Space Telescope data}

We also used the high-resolution images of the sample galaxies
available from the HST Archive. The HST images were obtained with
the optical WFPC2 and the infrared NICMOS cameras. WFPC2 has an
image scale of $0.1''/\mbox{px}$ and a field of view of 2\farcm7
and $0.045/\mbox{px}$ for the Planetary Camera (PC), with a field
of view of $36''$. The NICMOS detector includes three chips. Its
pixel scale and field of view are 0\farcs043 and $11''$ for NIC1,
0\farcs075 and $19''$ for NIC2, and 0\farcs2 and $51''$ for NIC3.
The information about the archival images is given in
Table~\ref{tab_hst}. Here ID is an archive identification of the
observational program.

\begin{table}
\caption{Log of direct imaging observations at the 6m telescope}
\label{tab_bta}
\begin{center}
\begin{tabular}{rrlr}\hline\hline
Name & Date & Filters & Seeing \\\hline
\object{NGC 470}  & 03/11/2000 & V, \Rc, SED755 & 1.4$''$\\
\object{NGC 2273} & 03/11/2000 & V, \Rc& 1.0$''$\\
\object{NGC 2681} & 25/10/2000 & V & 1.0$''$\\
\object{NGC 2950} & 04/11/2000 & V, \Rc & 1.5$''$\\
\object{NGC 3945} & 03/11/2000 & V, \Rc & 1.0$''$\\
\object{NGC 6951} & 04/11/2000 & V, \Rc, SED755 & 1.2$''$\\
\object{NGC 7743} & 03/11/2000 & V, \Rc, SED755 & 1.5$''$\\\hline
\end{tabular}
\end{center}
\end{table}

\begin{table} [t]
\begin{center}
\caption{HST high resolution direct images}
\label{tab_hst}
\begin{tabular}{rrrll}\hline\hline
Name & Date & Camera & Filters & ID\\\hline
\object{NGC 2273} & 05/02/1997 & WFPC2& F791W & 6419 \\
                  & 03/04/1998 & NIC2& F160W & 7172\\
                  & 03/04/1998 & NIC3& F164N, F166N & 7172\\
\object{NGC 2681} & 07/06/1998 & NIC3& F187N, F160W & 7919\\
                  & 03/09/2000 & WFPC2& F300W & 8632\\
\object{NGC 2950} & 26/01/1999 & WFPC2& F450W, F555W,& 6633\\
                  &            &      & F814W\\
\object{NGC 3368} & 04/05/1998 & NIC2& F160W & 7331\\
                  & 08/05/1998 & NIC2& F110W & 7331\\
                  & 12/12/2000 & WFPC2& F814W & 8602\\
\object{NGC 3786} & 30/03/1995 & WFPC2& F606W & 5479\\
                  & 29/04/1998 & NIC1& F110W, F160W & 7867\\
\object{NGC 3945} & 11/05/1997 & WFPC2& F450W, F555W& 6633 \\
                  &            &      & F814W\\
\object{NGC 4736} & 02/07/2001 & WFPC2& F814W & 9042 \\
\object{NGC 5566} & 04/04/1999 & WFPC2& F606W & 6359\\
\object{NGC 6951} & 30/10/2000 & WFPC2& F814W & 8602 \\
                  & 18/11/2000 & WFPC2& F606W & 8597 \\
\object{NGC 7743} & 30/05/1994 & WFPC2& F547M & 5419 \\
                  & 11/10/1994 & WFPC2& F606W & 5479 \\
                  & 16/09/1997 & NIC2& F160W & 7330 \\\hline
\end{tabular}
\end{center}
\end{table}

\subsection{Long-slit observations of \object{NGC 2681}}

When all the data were reduced and analysed we concluded that the
photometric and MPFS data were not enough for understanding the
outer morphology in \object{NGC 2681}; first of all, the  $PA$  of the disk
was very uncertain (Sect.~\ref{sec_2681}). For this goal
additional observations of this galaxy were performed in April-May
of 2003 at the 6m telescope with the SCORPIO focal reducer in a
long-slit mode. The detector was the EEV42-40 $2048\times2048$
pixels CCD array. The spectrograph reciprocal dispersion was
$0.85$\AA/pixel, and the spectral resolution was about $6$\AA. The
slit of $1.2''\times5.3'$ was centered onto the nucleus under
three fixed orientations near possible minor (PA=$30^\circ$ and
$61^\circ$) and major (PA=$121^\circ$) axes. The log of the
observations is given in Table~\ref{tab_long}. Though the seeing
was $2.5\arcsec -3.5\arcsec$ on April 7, this  is not important
for the study of the kinematics of the outer regions.. The primary
reduction included bias subtraction, flat-fielding and cosmic-ray
hits removal. Then the individual spectra were sampled along the
slit with a bin of $0.6''$ at  distances $r<10"$ from the center
and with a bin of $2.5''$ at  larger distances. The spectrum of a
He-Ne-Ar lamp was used for wavelength calibration. The
line-of-sight velocity distributions along the slit were
constructed by means of a cross-correlation technique. The spectra
of the K0III stars observed in the same nights were used as
velocity templates. The errors of the velocity measurements are
from $10\km$ in the galactic center to $20-30\km$ at  distances of
$r=40-45''$.

\begin{table}
\caption{Log of long-slit observations } \label{tab_long}
\begin{center}
\begin{tabular}{rrrrr}
\hline\hline
Name& Date   & PA &$T_{exp}$, sec & Seeing \\
\hline
\object{NGC 2681} &07/04/2003 & 30$^\circ$ & $4\times600$  & 2.6$''$ \\
\object{NGC 2681} & 07/04/2003 & 61$^\circ$ & $3\times900$  & 3.5$''$ \\
\object{NGC 2681} &01/05/2003 & 121$^\circ$ & $3\times1200$  & 1.5$''$ \\
 \hline
\end{tabular}
\end{center}
\end{table}

\section{Analysis of the velocity fields and photometric data}

\label{method}

The isophote analysis of the images was performed with the FITELL
program written by V.V. Vlasyuk (SAO RAS) which uses the
well-known algorithm of Bender \& M\"{o}ellenhoff (\cite{bender}).
It yields a surface brightness value ($I$) and the orientation
parameters of elliptical isophotes: position angle $PA$, and
ellipticity $\epsilon=1-b/a$, where $a$ and $b$ are the semi-major
and semi-minor axis of ellipses respectively.

The velocity fields were analyzed by means of the ``tilted-ring''
method (Begeman, \cite{begeman}), with fixed systemic velocity and
position of the center as described by Moiseev \& Mustsevoi
(\cite{mm}). Here we briefly describe the method. The velocity
field is broken up into  elliptical rings, with a $1-1.5''$ width
aligned with the direction of the line of nodes. Using the
$\chi^2$ minimization of the deviations of the observational
points from the model, we calculate all 6 parameters
characterizing  pure circular rotation at a radius $R$ in the
galactic plane. They are: the coordinates of the rotation center
on the sky plane ($x_0$, $y_0$), the systemic velocity $V_{sys}$,
the position angle of the major axis $PA_{dyn}$ (dynamical
position angle), the galactic plane inclination to the
line-of-sight $i$, and the mean rotation velocity $V_{rot}$.
However, since all galaxies contain large-scale bars in the inner
and spiral arms in the outer regions, the ``circular'' velocity
fields in the observed galaxies must be distorted by non-circular
motions. The non-circular motions cause in the systematic errors
when determining the disk orientation parameters (see Appendix in
Lyakhovich et al., \cite{lyakh}; Barnes \& Sellwood,
\cite{barsel}; Fridman et al., \cite{monument}).

In such a situation it is convenient to fix some model parameters.
A good first approach is to assume $x_0,y_0, V_{sys}, \mbox{and}\,
i=const$, leaving only two free parameters for fitting in each
ring: $V_{rot}$ and \pak. The \pak deviation from the line of
nodes of the whole disk ($PA_0$) is used to characterize the type
of non-circular motions (oval distortion, polar disk, etc.). In
the case of the motions of gas clouds in a bar potential, the
observable \pak ceases to be aligned with the disk's line of nodes
(Chevalier \& Furenlid, \cite{chev}, Moiseev \& Mustsevoi,
\cite{mm}). A turn of the ``dynamical axis'' must occur in the
opposite direction, with respect to position angle of the inner
isophotes, or to the ``photometric axis'' (Moiseev \& Mustsevoi,
\cite{mm}). The same effects should be also observed in the
velocity field of the stars, as  is shown by numerical simulations
(Miller \& Smith, \cite{mil}; Vauterin \& Dejonghe, \cite{vd}),
and by analytical calculations of the stellar motions under the
effects of a triaxial gravitational potential (Monnet et al.,
\cite{monnet}). On the other hand, in the case of a pure circular
rotation in a warped (polar,  inclined) inner disk, the dynamical
axis must follow the photometric one.

In determining the set of optimal galactic disk orientation
parameters we were faced with some problems. For the systemic
velocity $V_{sys}$ we took its mean value over all radii. As a
first approach, for the center position $x_0,y_0$ we took the
coordinates of the symmetry center of the velocity field. If this
``dynamical center'' is at a distance less than $1-1.5''$ from the
center of the inner isophotes of the continuum image, then the
photometric center is accepted as the ultimate position of the
center. In \object{NGC 470} (Sect.~\ref{sec_470}), the difference between
the positions of the dynamical and photometric centers exceeded
the error of their measurement (the case of a lopsided galaxy).
For this galaxy we adopted the dynamical center as the center of
rotation.

The main problem is to accurately calculate the values of the
position angle of the line of nodes ($PA_0$) and the inclination
($i$). To use the averaged parameters of the outer isophotes is
not always correct for this, because the outer parts of the spiral
structure distort these isophotes in galaxies possessing
well-developed spiral arms. The choice of radius range for which
the average values of $PA$ and $\epsilon$ are calculated  is also
not obvious. In addition, some galaxies (\object{NGC 3368}, \object{NGC 3945}) have
ring-like structures at large distances from the center. If a ring
has a resonant origin, then it could be elliptical in the galactic
plane (Buta, \cite{buta96}) with uncertain ellipticity for an
observer. Finally, the errors in flat-fielding and
background-removing procedures (mainly for JH\K images) also
distort the shape of the low-brightness isophotes of the outer
regions of galaxies. Therefore, we used the ``spiral criterion'',
proposed by Fridman et al. (\cite{monument}), for $PA_0$ and $i$
determinations for the main part of our sample. This method is
based on  applying of a Fourier-series expansion to the surface
brightness distribution in the galactic plane $I(R,\varphi)$,
where $R$ is the radial distance, and $\varphi$ the azimuthal
angle:

\begin{equation}
I(R,\varphi)=A_0(R)+\sum_{m=1}^{N}A_m(R)\cos(m\,\varphi+\phi_m(R))
\end{equation}

\noindent $A_m$, and $\phi_m$ are the amplitude and phase of the
$m$-th harmonic, and $N=8$ to $12$ is the maximum number of the
terms. If a $m$-armed spiral structure appears on the image, then
the line of the maximum of the harmonic number $m$ should  also be
spiral. Clearly, in the bar region the line of the maximum of the
$m=2$ harmonic  must be straight and elongated along the position
angle of the bar. It appears that the shape of the spiral, traced
by the maximum of the harmonics, is very sensitive to the
selection of the orientation parameters. It can be shown that, if
for a two-armed galaxy the inclination $i$ is underestimated, then
the line of the maximum of the $m=2$ harmonic  at large radii will
be fixed near the direction of the line of nodes.

Similarly, if the adopted $PA_0$ differs from the ``true'' value,
then the line of the maximum of the $m=2$ harmonic will be
approximate to the direction of the ``true'' $PA_0$ in the outer
part of the disk. Our experience shows that for moderate
inclinations ($i=30-70^\circ$), the parameters of disk
orientations may be calculated with an uncertainty of a few
degrees. A more detailed description of this method is given in
Fridman et al. (\cite{monument}). In Table~\ref{tab_pai} we
present the $PA$ and $i$ for our sample galaxies. During the
search for the orientation parameters we tried to obtain some
consistency between all three methods (isophotal analysis,
analysis of the IFP velocity fields, and the ``spiral
criterion''). Nevertheless,  preference was given to the ``spiral
criterion''. The details of the orientation results are considered
in the discussion of individual galaxies.

\begin{table}
\caption{Parameters of disk orientation}
\begin{center}
\label{tab_pai}
\begin{tabular}{rllll}\hline\hline
Name & \multicolumn{2}{c}{$PA_0$, $(^\circ)$} & \multicolumn{2}{c}{$i$,
$(^\circ)$ } \\
\hline
 \object{NGC 470} & $157$&$\pm1$ & $55$&$\pm1$\\
\object{NGC 2273} & $58$&$\pm2$ & $50$&$\pm2$\\
\object{NGC 2681} & $148$&$\pm5$ & $25$&$\pm5$\\
\object{NGC 2950} & $119$&$\pm1$ & $50$&$\pm2$\\
\object{NGC 3368} & $135$&$\pm5$ & $48$&$\pm3$\\
\object{NGC 3786} & $247$&$\pm5$ & $61$&$\pm2$\\
\object{NGC 3945} & $157$&$\pm2$ & $52$&$\pm2$\\
\object{NGC 4736} & $298$&$\pm6$ &$31^1$& \\
\object{NGC 5566} & $210$&$\pm3$ & $64$&$\pm2$\\
\object{NGC 5850} & $335$&$\pm10$ & $37$&$\pm5$\\
\object{NGC 5905} & $130^2$& & $40^2$&\\
\object{NGC 6951} & $140$&$\pm6$ & $41$&$\pm4$\\
\object{NGC 7743} & $270$&$\pm5$ & $40$&$\pm3$\\\hline
\multicolumn{5}{l}{$^1$ Mulder (\cite{mulder})}\\
\multicolumn{5}{l}{$^2$ van Moorsel (\cite{van})}\\
\end{tabular}
\end{center}
\end{table}

\section{The results}

\label{atlas0}

\subsection{The Atlas}

\label{atlas}

The images and the results of the velocity field analysis are
presented in  Figs.~\ref{470m}-\ref{6951ifp}. The left panels of
Figs.~\ref{470m}-\ref{7743m} show the MPFS data. The top row
contains the continuum images, the velocity field of the stars,
and the map of the line-of-sight velocity dispersion ($\sigma_*$).
The image in the brightest emission line of the ionized gas, the
velocity field of this emission line and the map of the velocity
dispersion of the ionized gas ($\sigma_{gas}$) are given in the
middle row. For the last map the width of the instrumental contour
was taken into account. There are no ionized  maps for \object{NGC 2950},
\object{NGC 3368}, \object{NGC 4736}, and \object{NGC 5566}, because emission lines are
absent in the MPFS spectra of these galaxies. A gray scale is
drawn, in $\km$, for the velocity fields and for the velocity
dispersion maps; while it is in $\mbox{ergs}\, \mbox{s}^{-1} \,
\mbox{cm}^{-2}/\Box''$ for the continuum and emission line images,
except for \object{NGC 5850} in the [N~II] emission lines, where the scale
is in arbitrary counts. The cross marks the position of the
adopted center of rotation in all MPFS maps. At the bottom we show
the radial dependences of the rotation velocity $V_{rot}$ and the
position angle on the dynamical axis \pak for the stellar and
gaseous components. The dotted line marks the adopted position
angle of the line of nodes ($PA_0$).

Examples of the images and isophotal analysis results are shown in
the right panels of  Figs.~\ref{470m}-\ref{7743m}. The left column
contains the ground-based images with the largest field of view,
while the right column contains the same data for the
circumnuclear regions $(r<5-20'')$, observed with the HST. The
best-resolution ground-based images are presented for those
galaxies where there are no HST data. The logarithm-scaled images
with the corresponding isophotes are presented in the top row of
the panels. In the middle row we present the residual brightness
distributions, after the removal of elliptical isophotes  from the
top images. In this case, the gray scale is linear. The thick
dotted line shows the maximum of the $m=2$ harmonic for the
surface brightness, overlapped onto the ground-based images. At
the bottom, we plot the parameters of the isophotes: $PA$ and
ellipticity versus major semi-axis. The scatter of the
measurements, through the different filters, gives us some
information about the errors of the elliptical model. The
horizontal dotted line corresponds to the $PA_0$ value, and the
vertical dotted line (in the left plot) marks the radial range,
which is zoomed in on  the right-hand plots.

The results of the IFP observations are shown in
Figs.~~\ref{470ifp}-\ref{6951ifp}, which include the continuum and
\Ha or/and [N~II] emission line images. A square-root gray-scale
is used. The cross marks the dynamical center. Also shown is the
velocity field with isovelocities. The thick black contour
corresponds to the systemic velocity, while the thin contours
indicate line-of-sight velocities in steps of $\pm50$ $\km$.  The
last plot shows the rotation curve $(V_{rot})$ and the kinematic
position angles (\pak).  Here the dotted line marks the $PA_0$
value. For \object{NGC 3368}, \object{NGC 4736}, and \object{NGC 6951} we present data in the
\Ha and [N~II] emission lines.

It is necessary to note briefly the errors in the $V_{rot}$
determination. The \textit{formal} errors of the model usually are
$1-5\km$ because every  elliptical ring for   fitting contains
from several tens up to hundreds of points (the latter for  IFP
velocity fields). Nevertheless, the rotation curve, determined in
the frame of pure circular motions, will be affected by systematic
errors on the order of the spread in the observational velocities
around the average values at the given radius (Lyakhovich et al.,
\cite{lyakh}; Fridman et al., \cite{monument}). The error bars of
$V_{rot}(r)$ in our figures must be interpreted as the deviation
of the line-of-sight velocities from the models of circular
rotation in each ring; i.e. the mean value of non-circular
velocities. Therefore the ``real'' circular rotation curve should
coincide with our curves within the errors which are given in
Figs.~\ref{470ifp}-\ref{6951ifp}.

In the following subsections we will present the results for individual
sample galaxies.

\begin{figure*}
\centering
\caption{\textbf{(To be seen in landscape)}\object{NGC 470}. The
\textbf{left} panel shows the results  of the MPFS-observations.
Here the top row contains the continuum images, the velocity field
of the stars, and the map of their velocity dispersion. In the
middle row the emission line image, the velocity field of the
ionized gas and the map of the velocity dispersion of the ionized
gas are given. At the bottom  the radial dependences of the
rotation velocity and the $PA$ of the dynamical axis are shown.
The dotted line marks the line of nodes ($PA_0$). In the
\textbf{right} panel the  left column contains the large-scale
image, while the right column contains the image of the nuclear
region. The logarithm-scaled images with the corresponding
isophotes are presented in the top row of the panel. The residual
brightness distributions are plotted in the middle row. The thick
dotted line shows the maximum of the m=2 harmonic. At the bottom
the $PA$ and ellipticity of the isophotes  are plotted. The
horizontal dotted line corresponds to the $PA_0$ value, and the
vertical dotted line (in the left plot) marks the radial range,
which is zoomed in on  the right-hand plots.  } \label{470m}
\end{figure*}

\begin{figure*}
\caption{The same as Fig.~\ref{470m} for \object{NGC 2273} \textbf{(To be
seen in landscape)} } \label{2273m}
\end{figure*}

\begin{figure*}
\caption{The same as Fig.~\ref{470m} for \object{NGC 2681} \textbf{(To be
seen in landscape)} } \label{2681m}
\end{figure*}

\begin{figure*}
\caption{The same as Fig.~\ref{470m} for \object{NGC 2950} \textbf{(To be
seen in landscape)} } \label{2950m}
\end{figure*}

\begin{figure*}
\caption{The same as Fig.~\ref{470m} for \object{NGC 3368} \textbf{(To be
seen in landscape)} } \label{3368m}
\end{figure*}

\begin{figure*}
\caption{The same as Fig.~\ref{470m} for \object{NGC 3786} \textbf{(To be
seen in landscape)} } \label{3786m}
\end{figure*}

\begin{figure*}
\caption{The same as Fig.~\ref{470m} for \object{NGC 3945} \textbf{(To be
seen in landscape)} } \label{3945m}
\end{figure*}

\begin{figure*}
\caption{The same as Fig.~\ref{470m} for \object{NGC 4736} \textbf{(To be
seen in landscape)} } \label{4736m}
\end{figure*}

\begin{figure*}
\caption{The same as Fig.~\ref{470m} for \object{NGC 5566} \textbf{(To be
seen in landscape)} } \label{5566m}
\end{figure*}

\begin{figure*}
\caption{The same as Fig.~\ref{470m} for \object{NGC 5850} \textbf{(To be
seen in landscape)} } \label{5850m}
\end{figure*}

\begin{figure*}
\caption{The same as Fig.~\ref{470m} for \object{NGC 5905} \textbf{(To be
seen in landscape)} } \label{5905m}
\end{figure*}

\begin{figure*}
\caption{The same as Fig.~\ref{470m} for \object{NGC 6951} \textbf{(To be
seen in landscape)} } \label{6951m}
\end{figure*}

\begin{figure*}
\caption{The same as Fig.~\ref{470m} for \object{NGC 7743} \textbf{(To be
seen in landscape)} } \label{7743m}
\end{figure*}

\begin{figure*}
\caption{IFP observations  in the \Ha line. \textbf{left} -- \object{NGC 470}. At
the top  continuum and \Ha images plotted on a square-root scale. The
cross marks the dynamical center. At the bottom the velocity field with
isovelocities is shown. The thick black contour corresponds to the
systemic velocity, while the thin contours indicate line-of-sight
velocities in steps of $\pm50$ $\km$. The last plot shows the rotation
curve  and the kinematic position angles.  Here the dotted line marks the
$PA_0$ value. There is a ghost from a bright star  located in position
$(20'', 35'')$. \textbf{right} -- same for \object{NGC 2273}. \textbf{(To be seen
in landscape)}  } \label{470ifp}
\end{figure*}

\begin{figure*}
\caption{ \object{NGC 3368}. IFP observations in the [NII] \textbf{(left)} and \Ha
lines \textbf{(right)}.  The dotted line correspond to the Position Angle of
the gaseous disk of $170^\circ$ (see text). \textbf{(To be seen in landscape)} }
\label{3368ifp}
\end{figure*}

\begin{figure*}
\caption{IFP observations of \object{NGC 3945} in the [NII] line. } \label{3945ifp}
\end{figure*}

\begin{figure*}
\caption{\object{NGC 4736}. IFP observations in the [NII] \textbf{(left)} and \Ha lines
\textbf{(right)}. \textbf{(To be seen in landscape)} } \label{4736ifp}
\end{figure*}

\begin{figure*}
\caption{\object{NGC 6951}. IFP observations in the [NII] \textbf{(left)} and \Ha
lines \textbf{(right)}. \textbf{(To be seen in landscape)} }
\label{6951ifp}
\end{figure*}

\subsection{\object{NGC 470}}

\label{sec_470}

The adopted orientation parameters, $PA_0=149\pm3^\circ$,
$i=51\pm4^\circ$, have negligible differences from the
measurements of Garcia-Gomez \& Athanassoula (\cite{gar}).
However, the parameters from Table~\ref{tab_pai} allow us to
obtain a more smooth well-ordered spiral of the maximum of the
$m=2$ harmonic. In the circumnuclear region, $PA$ of the SED755
isophotes matches the NIR data of Friedli et al. (\cite{fri96}),
but it deviates by more than $25^\circ$ from the measurements in
the V and \Rc bands. This  can be related to the dust extinction,
which is more significant in the optical bands. The deviations of
 \pak from the line of nodes at $r=8-20''$ (from the IFP data,
see Fig.~\ref{470ifp}) are explained by us as an effect of the
large-scale (primary) bar on the ionized gas velocities. The turn
of \pak occurs in the opposite direction with respect to the inner
isophotes $PA$ (Fig.~\ref{470m}), in agreement with existing
models of the gas flowing within the bars (see
Sect.~\ref{method}). Also the MPFS data reveal that the dynamical
axis, for the stellar component, turns in the same direction as
the gaseous (in \Ha) one at $r>5''$.

The center of rotation (dynamical center) of the MPFS velocity
fields of stars and gas is shifted by $\sim4''$ ($\sim0.6$\, kpc),
in the NW direction, from the photometric center. The discrepancy
between the positions of both centers is considerably larger than
the measurement errors or a possible effect of dust absorption.
The difference in the locations of the centers is also observed in
the large-scale \Ha velocity field, which  confirms the result of
Emsellem (\cite{eric470}). Such a discrepancy in the photometric
and dynamical centers locations or between the centers of the
inner and outer isophotes (lopsided-structure) have been seen in
numerous galaxies (Richter \& Sancisi, \cite{lops_obs}). The
nature of the lopsidedness could be different at different radial
scales, from the circumnuclear disk precessing around a
super-massive black hole to development of a spiral azimuthal
disturbance $m=1$ mode in the large-scale stellar-gaseous disk
(Junquera \& Combes, \cite{lopside}; Emsellem et al.,
\cite{emsellem01}; Emsellem, \cite{eric470}).

Turnbull et al. (\cite{turn}) have proved that \object{NGC 474} shows signs of
interaction with \object{NGC 470}. They also noted that both galaxies are perhaps
embedded in a common HI-cloud. Hence, the excitation of the non-axisymmetrical
harmonics in \object{NGC 470} may be caused by a tidal interaction with the companion.

An alternative explanation can also be proposed for the lopsidedness  since the
distorted region is relative small and located inside the large-scale bar. It
is possible to suppose an abnormal development of the non-axisymmetrical
harmonics in the surface brightness during the secular evolution of a barred
galaxy. A similar shift between the isophotal center and the dynamical center
is observed in the central regions of some barred galaxies (Zasov \&
Khoperskov, \cite{zashop}).

Wozniak et al. (\cite{woz}) assumed that the isophotal turn at $r<7''$ was
connected with a secondary bar. However later  Friedli et al. (\cite{fri96})
have explained this feature by a triaxial bulge in the primary bar. This
explanation is problematic from our point of view, because the position angle
of the inner isophotes (at the $r=2-6''$) is close to the disk $PA_0$ value.
Also the amplitude of isophotal $PA$ deviations from $PA_0$ decreases sharply
when passing from short wavelengths (V) to longer ones (SED755). Such
behaviour of the isophotes can be explained by dust extinction. Moreover, the
apparent ellipticity of the inner isophotes $\epsilon=0.3-0.45$
(Fig.~\ref{470m}) corresponds to the projection of a circular disk onto the
sky plane (if problems with bulge and outer bar deprojection are  taken into
account). In Sect.~\ref{disks} we decompose the surface  brightness
distribution in \object{NGC 470} and justify the assumption. Thus we think that there
is a circumnuclear disk within the large-scale bar (see Sect.~\ref{disks} for
additional arguments). In the \Ha velocity field the turn of \pak occurs in
the same direction with respect to the $PA$ of isophotes at $r<5-7''$
(Fig.~\ref{470ifp}). Such a dynamical feature is also in agreement with
circular motions in the decoupled disk which may be slightly inclined to the
galactic plane (Sect.~\ref{method}). An asymmetrical $m=1$ harmonic is
developed in this disk, which also has a large concentration of  molecular CO
(Sofue et al., \cite{co93}), as well as of  ionized hydrogen (see the
\Ha-image on Fig.~\ref{470ifp}). The decoupling of the central disk may be
caused by its location at the ILR of the large-scale bar.

Vega Beltra\'n et al., (\cite{vega}) gave an estimation of the central
velocity dispersion of stars; $\sigma_*=56\pm31\km$ from their long-slit
observations of \object{NGC 470}. However, such a low value of $\sigma_*$ is doubtful,
as it is difficult to explain the fact that the rotation velocity (from our
MPFS data) of stars in the central region is smaller by $ \sim 50\km$ than the
gas velocity. Our measurements support a central velocity dispersion of $\sigma
_*\approx140-150\km $, which looks more realistic and agrees with estimations
of Hera\'udeau et al. (\cite {hera}). The disagreement with the Vega Beltra\'n
et al. (\cite{vega}) results may be due to a template mismatching problem, at
was already marked in their article.

In the \Ha velocity field, at distances of $r=12-20''$ to the East from the
center, we observe a region with  line-of-sight non-circular motions about
$50-60\km$. The profile of the \Ha emission line has a complex multicomponent
structure. Clearly, two or more systems of  gaseous clouds, with different
velocities along the line-of-sight, are present in this abnormal region. The
non-circular motions apparently are not related to the bar dynamics because
they are not symmetric relative to the galaxy center. We suppose an
extragalactic gas cloud falling into the galactic disk. Similar shapes of the
emission line profiles (at much larger distances from the center) were
observed by us in \object{NGC 1084} (Moiseev, \cite{mois00}).

\subsection{\object{NGC 2273} (\object{Mrk 620})}

\label{sec_2273}

Van Driel \& Buta (\cite{vanbuta}) have described a complex morphology of this
galaxy: a large-scale bar of $\sim50''$ in diameter and two systems of spiral
arms ending at distances 1\farcm1 and 1\farcm6 from the center. The influence
of this large-scale bar appears clearly  in the IFP velocity field in the \Ha
emission line (Fig.~\ref{470ifp}): the dynamical axis deviates by more than
$20^\circ$ from the line of nodes $PA_0$ at radii $r=3-45''$. The turn of \pak
occurs in the opposite direction with respect to the $PA$ of the isophotes at
these distances (Fig.~\ref{2273m}), in accordance with a common picture of gas
flows in barred galaxies. The residual line-of-sight velocities along the
minor axis have amplitudes of more than $50-80\km$ (after pure circular model
subtraction) at  radial distances of $\pm 10-13''$. The sign of the residual
velocities corresponds to the gas inflow motions\footnote{We assume that
spiral arms are trailing. Thus the NW-side of the galaxy is the nearest to an
observer.} due to the large-scale bar. The \pak radial variations in the \Hb
velocity field (from the MPFS observations) are in agreement with those in the
\Ha velocity field. Nevertheless, the amplitude of the \Hb rotation curve (at
$r=1-6''$) is less by $\sim50\km$ than the  \Ha rotation curve, and almost
coincides with the rotation velocity of stars (Fig.~\ref{2273m}). Apparently,
the relatively weak \Hb emission line is distorted by the broad absorption
line from the underlying stellar spectrum.

The [OIII] velocity field is significantly different  from the measurements in
the Balmer lines. At $r<3''$ the [OIII] line-of-sight velocities are constant,
therefore the \pak and $V_{rot}$ are uncertain in this region. At  larger
distances from the center the [OIII] gradient of the line-of-sight velocities
is smaller than that of the hydrogen lines. We suggest that the existence of
non-circular motions in forbidden lines is caused by the jet propagating
outward from the Seyfert 2 nucleus. Ferruit et al. (\cite{ferruit}) found a
jet-like feature in the HST [OIII]-images and ionized maps
($\mbox{[OIII]}/\mbox{H}_\alpha$). It is seen, up to $\sim2''$ to the east
from the nucleus and coincides with the extended radiostructure. The spatial
geometry of the gas motions around the jet cannot be restored from the MPFS
data, most probably these motions are outside the galactic plane.

The stellar velocities do not reveal any non-circular motions
in contrast with the gaseous velocities. The elliptical structure,
elongated in E-W direction, appears in the stellar velocity
dispersion map with a small drop of $\sigma_*$ by $10-15\km$ in the
rotation center (Fig.~\ref{2273m}).

Mulchaey et al. (\cite{mulch}) have discussed a secondary bar at $r<8''$ from
their NIR-photometry data. The high-resolution HST images showed that this
``bar'' breaks up into several elliptical arcs with radii of $r=3-4''$ (see the
residual brightness image in Fig.~\ref{2273m}). Ferruit et al.
(\cite{ferruit}) described these features as a ``pseudo-ring'', and we suggest
that it is a symmetrical spiral twisted into a ring. This conclusion agrees
with the opinion of Erwin \& Sparke (\cite{erw02}, \cite{erw03}). However
Petitpas \& Wilson (\cite{petit}) returned toward a secondary bar assumption
in the central kpc of \object{NGC 2273} based on their radio observation. We will
discuss the molecular gas properties of the sample galaxies in Sect.~\ref{co}.
It will be shown that in the case of \object{NGC 2273} the CO morphology agrees with an
assumption of the inner stellar-gaseous disk embedded into the large-scale
bar. The disk lies in the galactic plane and possesses a circumnuclear spiral,
as  is mentioned above.

\subsection{\object{NGC 2681}}

\label{sec_2681}

The sharp variations of the isophotal parameters (Fig.~\ref{2681m}) reflect the
complex morphological structure of the galaxy. Friedli et al. (\cite{fri96})
connected the radial turns of the isophotes with two bars, first proposed by
Wozniak et al. (\cite{woz}). Erwin \& Sparke (\cite{erw99}) offered a model of
the circumnuclear structure of \object{NGC 2681}, consisting of three independent bars,
with position angles of the major axes $PA_1=30^\circ$ $(20''<r<60''$),
$PA_2=75^\circ$ $(4''<r<20'')$ and $PA_3=20^\circ$ $(r<4'')$. The smallest
bar-like structure is  seen in the HST NICMOS images (Fig.~\ref{2681m}).

 However, only  isophotal analysis results are not enough  to
prove triple-barred structure because the photometry
of the galaxy is a puzzle.  We must note two items.

 Firstly, this lenticular  galaxy has no global spiral arms but its optical
color maps (Wozniak et al., \cite{woz}; Erwin \& Sparke, \cite{erw99}) reveal
short tightly  wound spiral arms and numerous dust lanes in the region of the
``middle bar'' ($r=10-20''$).  The majority of the dust lanes are located to
the south of the nucleus and are extended along the spiral arms. These
flocculent mini-spirals also can be seen in our model-subtracted V-band  image
(Fig.~\ref{2Dprf}) and unsharp masked R-band image by Erwin \& Sparke
(\cite{erw99}).

 Secondly, the disk's position
angle is uncertain from available imaging data. It varies from $PA_0=50^\circ$
(Friedli et al., \cite{fri96}) to $130-140^\circ$ (Erwin \&
Sparke,\cite{erw99}; \cite{erw03}). The outer V-band isophotes
(Fig.~\ref{2681m}) have $PA=110-120^\circ$.  Also the  \pak in the
circumnuclear region (Fig.~\ref{2681m}) differs  from all these values. We
tried to use a Fourier expansion method to search for disk orientation
parameters and failed completely, because there are no symmetrical spirals in
the galaxy.

\begin{figure}
\resizebox{\hsize}{!}{\includegraphics{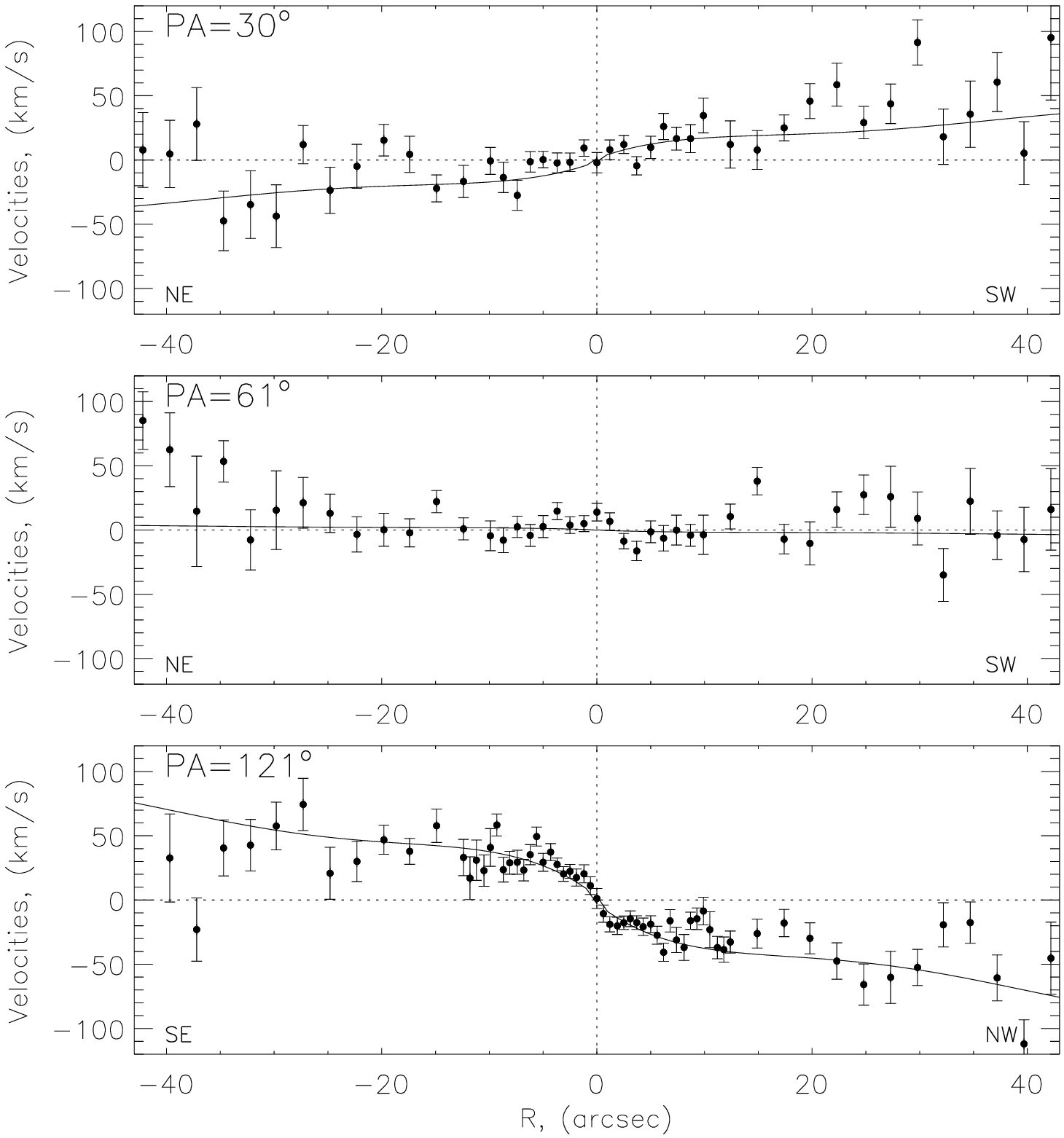}}
\caption{ \object{NGC 2681}: Stellar velocities  measured along different position
angles. The solid line marks the best-fit  model of the rotation curve}
\label{2681long}
\end{figure}

The long-slit  data help to determine the disk orientation.
Fig.~\ref{2681long} shows that the line-of-sight velocity gradient is absent
along  $PA=61^\circ$. Hence this is the direction of the minor axis (here we
neglect possible non-circular motions in the bars). For more precise analysis
we fitted velocities in all cross-sections by a model of pure circular
rotation. The following values were found from a $\chi^2$-minimization:
$PA_0=148\pm5^\circ$ and $i=25\pm5^\circ$. These parameters roughly agreed
with the orientation of the outermost isophotes $PA_0=140^\circ$ and
$i=18^\circ$ (Erwin \& Sparke, \cite{erw03},  they do not quote an
uncertainty, but it is probably at least $3^\circ$).

The panoramic spectroscopy data (Fig.~\ref{2681m}) show that stars at
$r>3-4''$ exhibit a circular rotation where the value of \pak matches the
disk's line of nodes. But at $r<2''$ we observe strong changes of the
dynamical axis of the stars  by $30^\circ-90^\circ$. The \pak variations in
the velocity field of the ionized gas are smaller, but differ systematically
from that of the stars  by $10-15^\circ$ at $r>3''$. Also the symmetry center
of the gas velocity field is shifted by $1.5 ''$ to the east from the
continuum center, similarly to \object{NGC 470}.  In addition, a sharp turn of the
isovelocities is seen at $r<2''$ (80 pc). So this region is dynamically
decoupled concerning both the stars and the gas.  At these distances, the
value of \pak of the stars is closer to the orientation of the HST inner
isophotes ($PA=25-35^\circ$). Such a relation between the ``photometric'' and
``kinematic'' axes  does not point in the direction of the bar hypothesis, but
of  disk-like kinematics. Thus an inclined (or polar) stellar-gaseous disk
exists in the central few tens parsecs of \object{NGC 2681}. This disk is seen on
HST-images of the galaxy from UV (WFPC2) to NIR (NICMOS). Erwin \& Sparke
(\cite{erw99}) accepted this disk as a third (inner) bar on HST-images.
Unfortunately, the spatial sampling of the MPFS data is not sufficient for
more certain conclusions on kinematics in this region (the seeing was $1.7"$).

 Cappellari et al. (\cite{cap}) have proved that \object{NGC 2681} hosts a nuclear
starburst of $\sim 10^9$ yrs old. They considered a merger process with a
small gaseous companion as the most probable trigger of the nuclear star
formation. If the companion had a spin direction inclined to the galactic
plane, then the remnants of the merged gas will be observed as a central
inclined disk, and the non-orthogonal disk will  precess toward the galactic
plane. However, a polar disk may be stable enough and rather long-living
(Arnaboldi \& Sparke, \cite{polar_t}). From this point of view the central
structure of \object{NGC 2681} is similar to that of galaxies with inner polar rings,
such as IC~1689 (Hagen-Thorn \& Reshetnikov, \cite{hagen}) or UGC~5600
(Shalyapina et al., \cite{u5600}). Nevertheless, the polar disk in \object{NGC 2681} is
smaller, of about $ 100-300$~pc in diameter.

The merger hypothesis makes it possible to explain the spiral arms and dust
lanes at $r<20''$, which have a shape differing from ordinary dust lanes in
large-scale spiral arms. Also these flocculent non-symmetrical spirals differ
from  the symmetrical two-armed dust spirals which are usually observed near a
shock front in barred galaxies. The \Ha$+$[NII] ionized gas emission lines is
also observed only at $r<15''$ (Wozniak et al., \cite{woz}). Thus, the
stellar-gaseous disk, with spiral arms and perturbed dust layer\footnote{We
suggest that the spiral in \object{NGC 2681} probably has a dust origin. Indeed, the
spiral appears in the optical images and is absent in  the J-band residual
image (Fig.~\ref{2681m}) as well as in  NIR color maps (Friedli et al.,
\cite{fri96})}, is nested in the center of the large scale-bar at $r<15-20''$
($600-900$~pc). A strong asymmetry in the dust distribution may be associated
with a possible warp of the gaseous-dust disk relative to the stellar one. We
suggest that all the features mentioned above (inner polar disk, abnormal
shape of the spirals, and warp of dust layer) are caused by a recent merging
event.

\begin{figure}
\resizebox{\hsize}{!}{\includegraphics{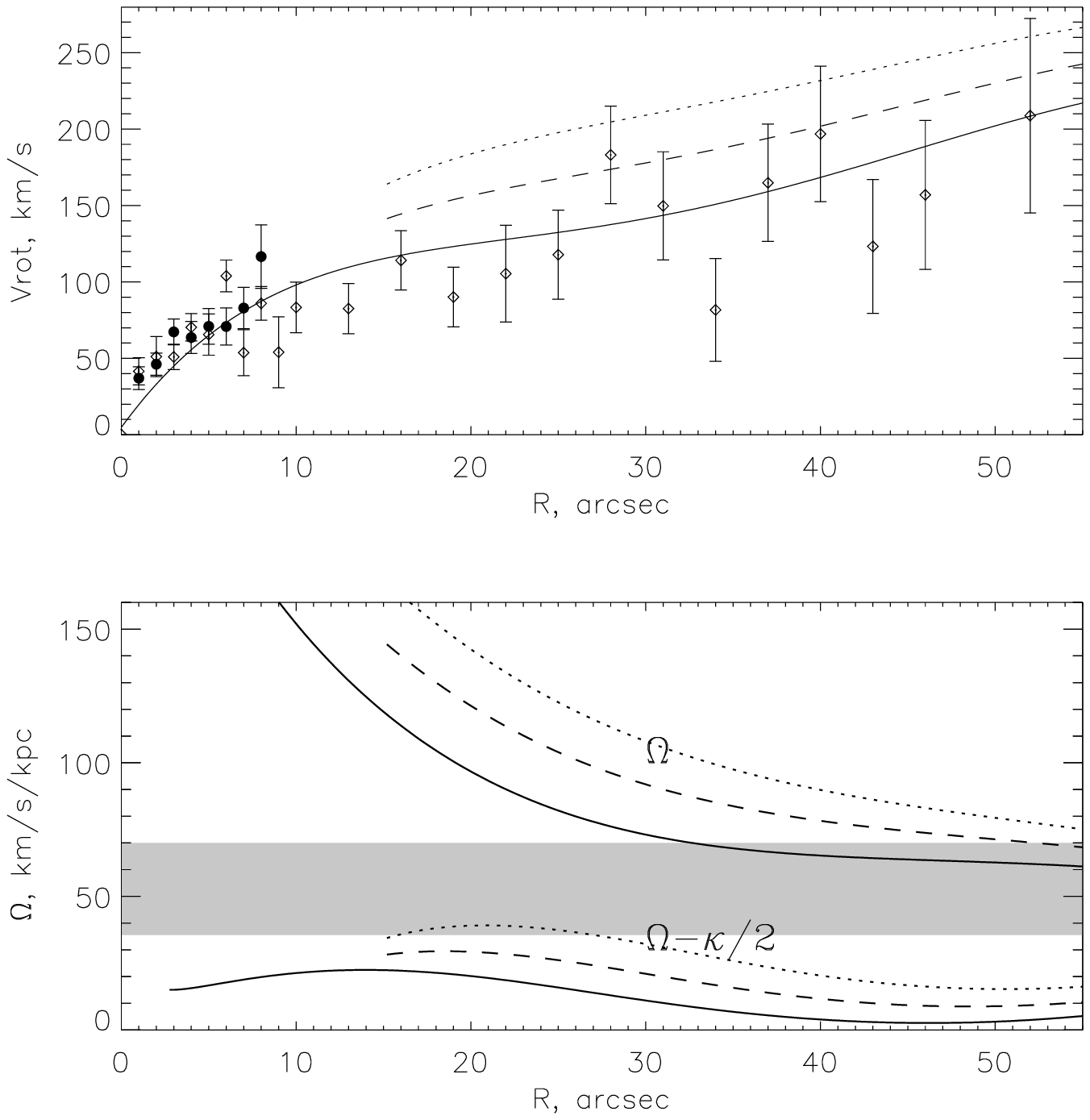}}
\caption{\textbf{top} Rotation velocities  of the stars in \object{NGC 2681}.
Different symbols show MPFS (filled) and long-slit (open) data. The solid
line is a polynomial fitting. The other lines mark the velocity streaming
curves where asymmetric drift terms are taken in account with
$\sigma_z/\sigma_R=0.7$ (dotted) and $1.0$ (dashed). \textbf{bottom} The
angular velocity ($\Omega$) and $\Omega-\kappa/2$ (where $\kappa$ is the
epicyclic frequency) curves calculated from the different models of the
rotation curve. The thick rectangle shows the possible range of the
angular velocity of the bar.}
 \label{2681ilr}
\end{figure}

We try to estimate the positions of ILRs in \object{NGC 2681}. Figure~\ref{2681ilr}
shows the polynomial approximation of the stellar rotation curve. However, the
amplitude of circular velocities must be larger than these streaming
velocities because velocity dispersion provides additional support against
gravity. For asymmetric drift correction we used Eq.~(2) from Aguerri et al.
(\cite{fast}) and their assumptions (the ratio of azimuthal to radial
components of the velocity dispersion $\sigma_\phi/\sigma_R=1/\sqrt(2)$,
$\sigma_*$ and a decrease of the volume density of the disk that is exponential
with radius). The drift correction was calculated for $\sigma_Z/\sigma_R$
values 0.7 and 1.0, bracketing the possible range for early-type galaxies
(Aguerri et al., \cite{fast}). The corrected circular velocities and  the
resonance curves shown in Fig.~\ref{2681ilr} are truncated at the radius of
$15''$ because the asymmetric drift approximation fails in the bulge-dominated
region (see surface brightness profile in Fig.~\ref{2Dprf}). For the
calculation the angular velocity $\Omega_P$ of the bar we make the following
assumptions: the maximal circular velocity is about $220-260\km$; the bar size
is $R_b=60''$; the ratio of the radii can be 1 to 1.6 $R_b$, this is the range
observed for bars in lenticular galaxies (Aguerri et al., \cite{fast}). The
probable range of the large-scale bar angular velocity,
$\Omega_P=35-70\km\,\mbox{kpc}^{-1}$, is marked as gray rectangular in
Fig.~\ref{2681ilr}. This figure also demonstrates that one or two ILRs could
exist near the radius  $r\approx20''$, if the $\Omega_P$ is smaller than the
value  mention above. This distance coincides with the approximate size of the
inner dust/gaseous disk described above. In this case the gas was accumulated
by the gravitational forces of the bar. Of course we must remember that the
resonance curves are calculated for a symmetric potential without any bar
perturbations. But we think that our evaluation of the ILR radius can be used
as a zero approximation of the position of the resonances.

In conclusion, we suggest the following interpretation of the global
morphology of  \object{NGC 2681}. The large-scale low-contrast bar dominates in the
optical and NIR images at $r=50-60''$ and has the ILRs at $r\approx15-25''$.
The decoupled gaseous disk is observed inside the resonance region. This inner
disk is elliptical (non-circular) because the elongated orbits of stars change
their major axis orientation at the resonances (Lindblad, \cite{lind}). This
elliptical disk was accepted by Erwin \& Sparke (\cite{erw99}) as a secondary
(``middle'') bar. The recent dwarf companion merging leads to the structure
perturbation of the kiloparsec-size disk with inclined (polar) orbits in the
central 100 pc.

\subsection{\object{NGC 2950}}

\label{sec_2950}

Our value of $PA_0$ differs slightly (by $10^\circ$) from the measurements of
Friedli et al. (\cite{fri96}). The radial dependences of the isophotal $PA$
and $\epsilon$, in all observed bands, coincide with those of Wozniak et al.
(\cite{woz}) and Friedli et al. (\cite{fri96}). The large-scale bar
($r\approx30-40''$ in size) is seen over the distributions of the $m=2$
Fourier harmonic and has a turn of isophotal $PA$ accompanied by an
ellipticity peak (Fig.~\ref{2950m}).

Wozniak et al. (\cite{woz}) and Friedli et al. (\cite{fri96}) explained the
inner $PA$ turn, and the peak, at $r<6''$, as a secondary bar. Surprisingly,
this ``bar'' does not distort the circular rotation of stars in the
circumnuclear region. The \pak of the stellar velocity field coincides with
$PA_0$ of the line of nodes.

We do not detect any non-circular motions in this velocity field despite the
strong turn of isophotes at these radii. There are sharp elliptical structures
decoupled in the stellar velocity dispersion map at $r=1-5''$
(Fig.~\ref{2950m}). It is extended along $PA\approx150^\circ$, which is the
direction of the major axis of the primary (external) bar.

\subsection{\object{NGC 3368} (Messier~96)}

\label{sec_3368}

Sil'chenko et al. (\cite{leo}) analyzed  the results of panoramic spectroscopy
of this galaxy including our observational data. However, that work is mostly
focused on the stellar population properties. We will briefly describe the
main properties of the inner kinematics of this galaxy, which is a member of
the Leo~I group.

Observations with the scanning IFP allow us to map the large-scale velocity
fields of the gas in two emission lines (Fig.~\ref{3368ifp}). The \Ha emission
is mostly concentrated in the starforming ring, at distances of $50-70''$ from
the center. Unfortunately, the stellar absorption features and the overlapping
interference orders  distort the emission line profiles in the \Ha data-cube,
over the whole central region of the galaxy. We cannot resolve this problem
because the free spectral range of the IFP is rather small (28\AA). In
contrast, the [NII] velocity field has been constructed over the full range of
radii. The dynamical position angle of the outer gaseous disk,
$PA_{dyn}=170^\circ$, at $r=30-200''$, differs strongly from the orientation
of the line of nodes of the stellar disk, $PA_0=135^\circ$. The last value was
calculated from the Fourier expansion of the NIR images (``spiral criterion'',
Sect.~\ref{method}).  So the line of the maximum of the $m=2$ harmonic has a
regular spiral shape and fits both spiral arms of the residual brightness
distribution (Fig.~\ref{3368m}) only for this value of $PA_0$. The isophotes,
distorted by these spiral arms, were erroneously interpreted by Jungwiert et
al. (\cite{jung}) as a large-scale primary bar.

The orientation of the ionized gas disk $PA_{dyn}=170-175^\circ$ is in  good
agreement with the outer part of the neutral hydrogen velocity field
(Schneider, \cite{hi3368}) and with CO kinematics at $r<15''$ (see Fig.~1 in
Sakamoto et al., \cite{sakamoto}). Thus, we suggest that the gas
(ionized/neutral) and the stars in this galaxy rotate in different planes. All
the gas in this galaxy, which is the brightest member of the western half of
Leo group, has an external origin, in agreement with Schneider (\cite{hi3368}).
See Sil'chenko et al. (\cite{leo}) for detailed arguments in favor of this
possibility.

Good agreement exists between the measurements of the \pak from two independent
IFP velocity fields. Nevertheless, the [NII] rotation velocities are
systematically  lower, by $\sim50\km$, than the \Ha ones. If the gaseous disk
is inclined to the main galaxy plane, then shock-wave fronts can  develope at
the cross-section of the global stellar and gaseous disks, because the gas
strikes the gravitational well. Thus, the low [NII] velocities may be
explained if the shock-excited gas, which is slowed down by collisions with
the stellar disk and spiral arms (these intensify the contrast of the
gravitation potential), emits mostly in the forbidden emission line. The
discrepancy between the velocities in the different emission lines is not an
artificial instrumental effect. \object{NGC 4736} was also observed, with the same
technique, in \Ha and [NII] lines during the same observation run. In the
\object{NGC 4736} data cubes the velocities in both emission lines are consistent with
each other.

There is no large-scale  (``primary'') bar  in this galaxy, as it was
mentioned above. However, a small-scale ($r\approx5''$) nuclear mini-bar is
present and extends along $PA=120^\circ-125^\circ$. This bar provokes a turn
of \pak in the stellar velocity field by $\sim25^\circ$, from the line of
nodes in the opposite direction with respect to the $PA$ turn of the inner
isophotes (Fig.~\ref{3368m}). The central asymmetry of the $\sigma_*$
distribution is confirmed by long-slit spectroscopy of Vega Beltra\'n et al.
(\cite{vega}).

The sharp turn of the isophotes, in the HST images inside $r=2''$, may be
associated with strong dust lanes. Each of these lanes crosses the images near
the nucleus and can be seen both in the optical and NIR images
(Fig.~\ref{3368m}). In Sil'chenko et al. (\cite{leo}) we argue that a
gaseous-dust mini-disk is located in the polar plane of the bar.
Unfortunately, our MPFS and IFP data have too low angular resolution to study
the polar disk kinematics.

\subsection{\object{NGC 3786} (Mrk~744)}

\label{sec_3786}

Our disk orientation parameters for this galaxy are in agreement with the
results of  Afanasiev \& Shapovalova (\cite{af_shap81}). Afanasiev et al.
(\cite{trans}) from their imaging data found two bars with semi-major axes of
$7''$ and $25''$ in this Sy~1.8 galaxy. However, our Fourier analysis of the
optical and NIR images shows that the line of the maximum of the $m=2$
harmonic, at $r>6-7''$, is wound into a regular spiral that coincides with the
global two-armed spiral pattern of the galactic disk. This spiral also appears
in the residual brightness maps (Fig.~\ref{3786m}). Thus, the radial
variations of $PA$ of the inner isophotes caused by the spiral arms have led
Afanasiev et al. (\cite{trans}) to an erroneous conclusion about a large-scale
bar in \object{NGC 3786}. The situation is the same as in the case of \object{NGC 3368}.

The \pak of the velocity field of stars at $r<6''$, differs from the  $PA_0$
of the line of nodes by more than $10^\circ$. The deviations occur in the
opposite direction relative to the central isophotes (Fig.~\ref{3786m}). This
indicates a dynamically decoupled central mini-bar of $2$~kpc in diameter. The
line of the maximum of the $m=2$ harmonic at $r < 5-6''$ has a constant
azimuthal angle, a further confirmation of the existence of the mini-bar. The
mini-bar was predicted by Afanasiev \& Shapovalova (\cite{af_shap81}) from
their long-slit data and confirmed photometrically by Afanasiev et al.
(\cite{trans}).

The ionized-gas velocity fields in the \Hb and [OIII] lines are different.
More precisely, the  orientations of \pak in velocity field of the stars and
gas in \Hb coincide, but the position angle, calculated from the [OIII]
velocity field of the gas, deviates from the stellar one. The deviations may
be due to the presence of shock fronts at the bar edges, formed in the disk
gas under the action of the gravitational well of the bar. The post-shock gas
decelerates and emits in forbidden lines (Afanasiev \& Shapovalova
\cite{af_shap81}, \cite{af_shap96}). The rotation velocities in the [OIII]
line are smaller than those in the \Hb (Fig.~\ref{3786m}). This feature may
also be connected to braking shock-excited gas at the bar edges.

The  stellar velocity dispersion field has   drops  by $\sim 50\km$ in the
center  compared with $\sigma_*$ at $r=3-4''$. This feature will be discussed
in  Sect.~\ref{dispdis}.

\subsection{\object{NGC 3945} }

\label{sec_3945}

The radial variations of the isophotal parameters over all observed bands are
in accordance with early observations of Friedli et al. (\cite{fri96}) and
Erwin \& Sparke (\cite{erw99}). The global spiral pattern is absent in this
galaxy. The parameters of the disk orientation were calculated under the
assumption of a circular shape of the outer low-brightness ring
($r=120-150''$). The values of $PA_0$ and $i_0$ agree (within the errors given
in Table~\ref{tab_pai}) with the results of Erwin \& Sparke (\cite{erw99}) and
Erwin et al. (\cite{erwet03}).

A large-scale bar extends along the galactic minor axis. Erwin \& Sparke
(\cite{erw99}) accepted that the circumnuclear disk nests in the bar, because
the isophotal $PA$ matches $PA_0$ at $r=4-18''$. They also favored  the
existence of a secondary small bar embedded in the disk, and  found a
low-contrast ring at $r=7''$ as well. This ring is seen in our residual
HST-images (Fig.~\ref{3945m}). The origin of this ring is discussed by Erwin
et al. (\cite{erw01}).

The velocity field of the stars in the circumnuclear disk shows a regular
circular rotation (\pak coincides with the $PA_0$). That the kinematics is
disk-like agrees with the conclusion by Erwin et al. (\cite{erwet03}) that the
inner disk is more luminous (and massive)  than the bulge in this  region.
Also we find no any non-circular motions of the stars due to a  possible
secondary bar at $r<3''$ suggested  by Erwin \& Sparke (\cite{erw99}).  The
rotation curve agrees with early measurements of Bertola et al.
(\cite{bertola95}). A small difference in the amplitude of the rotation curves,
observed in 2000 and 2002, is probably due to a velocity gradient, smoothed by
beaming effects, in observations with a worse seeing in 2000
(Table~\ref{tab_mpfs}).

The velocity field of the ionized gas is puzzle. Bertola et al.
(\cite{bertola95}) found  [OII]$\lambda3727$\AA\, line emission only in the
very central region ($r<3''$) and showed that the gas co-rotates with the
stars. Nevertheless, our MPFS data contradict their conclusions. First of all,
we confidently detect the [NII] line emission up to a distance of $10''$ from
the center. Secondly, inside the radius of $6''$ (about $0.5$~kpc), the
line-of-sight velocities of the gas are inverted with respect to the stellar
velocities. Thus, a counter-rotating gaseous disk is located at $r<5-6''$.

The direction of gas rotation becomes coterminous with the stellar rotation at
 larger distances from the nucleus (see Fig.~\ref{3945m}). The large-scale
IFP velocity field (Fig.~\ref{3945ifp}) confirms the fact of normal gas
rotation at large radii, up to $140''$ ($\sim11$~kpc). The variations of \pak
at $r=5''-35''$ in the IFP velocity field are not real. This artifacts are
caused by spectral overlapping of the strong stellar \Ha absorption and
low-contrast [NII] emission lines in the central region. The small spectral
range of the IFP observations does not allow us to separate absorption and
emission lines. In contrast, at $r=115''-140''$ the stellar continuum has a low
contrast, while the [NII] lines are bright since they are emitted by some
HII-regions in two spirals twisted into the outer ring.

\subsection{\object{NGC 4736} (Messier~94) }

\label{sec_4736}

The outermost isophotes of our NIR images are limited to $r\approx50''$, and
we cannot calculate the parameters of the outer disk orientation from these
images. The application of the ``tilted-ring'' model to the [NII] velocity
field at $r=80-120''$ makes it possible to obtain the $PA_0$ value. The disk
inclination is small and uncertain from velocity field analysis. Mulder
(\cite{mulder}) found $i_0=31^\circ$ from the outer optical isophotes. We
accept this value throughout our paper.

The stellar velocity field shows a regular circular rotation with
$PA_{dyn}=300-305^\circ$ (Fig.~\ref{4736m}). These values differ slightly from
the  accepted $PA_0$, but fall inside the $PA_0$ errors (Table~\ref{tab_pai}).
From long-slit observations, Mulder (\cite{mulder}) obtained
$PA_{dyn}=285^\circ-290^\circ$ for the stellar component in the galactic disk
as a whole. This value differs certainly from our measurements at $r<10''$.

Our \Ha velocity field is similar to the data of Mulder (\cite{mulder}), but
it has no measurements inside $r<25''$. Here the broad \Ha absorption line
distorts the emission line profile. The absorption does not affect the [NII]
emission. The [NII] velocity field reveals non-circular ionized gas motions,
as is shown from the fact that \pak turns at $r=5-30''$. The non-circular
motions are most likely connected  with bar-like structure at $r<15-20''$,
detected from isophotal parameters variations: a turn of $PA$ and a $\epsilon$
peak (Fig.~\ref{4736m}). This conclusion confirms the result by Wong \& Blitz
(\cite{wong}) who suggest that molecular gas kinematics is consistent with
radial flows in the  central stellar bar ($\sim30''$ in diameter).

 Shaw et al. (\cite{shaw}) also considered a twist of NIR isophotes at
$r<5-7''$, which may be explained as a secondary bar effect. We must notice
that the most highest-contrast part of the circumnuclear mini-spiral is located
at these radii and distorts the shape of the isophotes in the HST images. This
spiral structure is studied in  recent work by Elmegreen et al.
(\cite{elmegreen}).

The dust lanes in the spiral appear in the HST image up to $r=20-30''$. Thus,
the deviations of \pak of the ionized gas velocity field from $PA_0$ at these
radii can be explained by  non-circular gas motions in the mini-spirals
instead of motions inside a large-scale  bar as was mentioned above.

\subsection{\object{NGC 5566} }

\label{sec_5566}

The disk inclination is considerably smaller than the value of $i_0=80^\circ$
measured by Jungwiert et al. (\cite{jung}). We found $i_0=64^\circ$ by means
of the ``spiral criterion''. This value is in  good accordance with the outer
isophote ellipticity (Fig.~\ref{5566m}) and with Kent's (\cite{kent})
CCD-observations. The large-scale bar of $25''$ in size is elongated at
$PA\approx155^\circ$. Two strong spiral arms start from the bar ends. The
velocity field of stars lacks significant non-circular motions as \pak
coincides with $PA_0$ within the errors. The stellar rotation velocity at
$r=10-14''$ reaches $250\km$, which corresponds to the HI rotation curve of
Helou et al. (\cite{helou}).

Jungwiert et al. (\cite{jung}) explained the isophotal ellipticity peak and
the $PA$ turn at $r<5''$ as secondary bar features. We disagree with this
interpretation for the following reasons. First of all, the position angles of
the ``secondary bar'' and the line of nodes are very similar. Secondly, two
symmetrically twisted spiral arms appear in the JH\K residual images. This
distribution of the underlying   stellar component is not characteristic for a
bar (Fig.~\ref{5566m}). And finally, a grand-design two-armed nuclear spiral is
seen also in the HST residual image (Fig.~\ref{5566m}) at these radii.

We must note that a slowly rotating  mini-bar can produce a spiral-like
response in the disk \textit{outside} the ends of the bar, where an ILR must be
located, as  has been shown in the numerical simulations by Combes \& Gerin
(\cite{comb}) and in the analytic estimations  by Fridman \& Khoruzhii
(\cite{fridman}). But in the case of \object{NGC 5566} the mini-spiral develops
\textit{inside} the radius, where a possible nuclear mini-bar is proposed by
Jungwiert et al. (\cite{jung}). Secondly, we find no non-circular motions
produced by a possible nuclear bar in the stellar velocity field, because \pak
agrees with the disk $PA_0$ (see above). And thirdly, a possible bar-like
structure is not visible  in the residual brightnesses deprojected to the
galactic plane (see Fig.~\ref{2Dprf}).

For all these reasons we suggest that the mini-spiral is  not connected with a
possible secondary bar.  Therefore,  \object{NGC 5566} is an ordinary barred galaxy. It
has a circumnuclear disk ($r=5''-8''$, or $500-750$ pc in linear radius) with a
mini-spiral nesting inside this bar. The properties of the inner isophotes
($PA=205-210^\circ$, $\epsilon_{max}=0.55$) indicate that this kiloparsec-size
inner disk lies in the galactic plane. The mini-spiral is traced by  dust
lanes but the majority part of the dust clouds is concentrated outside the
spiral arms in the NW half of HST PC2 image.

\subsection{\object{NGC 5850}}

\label{sec_5850}

Since our NIR images have a field of view that is  not large enough to cover
the entire galaxy, we used the R-band image, from the digitized atlas of Frei
et al. (\cite{frei}) to determine the orientation parameters. The spiral arms
are low-contrast outside the ring around a large-scale bar ($r>80''$). Higdon
et al. (\cite{higdon}) have shown that the outer spirals have a mixed (2- and
3-armed) structure. However, the orientation parameters, determined by means
of the ``spiral criterion'', from the $m=2$ harmonic are in good agreement
with preliminary data of Friedli et al. (\cite{fri96}), and with the kinematics
of neutral hydrogen if the possible warp of the HI-disk is taking into account
(Higdon et al., \cite{higdon}). Since the inner isophotes have a sharp turn at
$r<10''$, Friedli et al. (\cite{fri96}) have assumed the existence of a
secondary bar in this galaxy.

The MPFS data show that \pak of the stellar rotation coincides with $PA_0$,
while in the ionized gas \pak, in the circumnuclear region, differs  by more
than $50-60^\circ$  from the line of nodes position angle (Fig.~\ref{5850m}).
Note that the ionized gas \pak almost aligns with the $PA$ of the inner
isophotes. This radial behavior of \pak is typical for a disk inclined to the
galactic plane. Nevertheless, if we assume that the gas motions are produced
in the galactic plane, they would correspond to a radial outflow from the
nucleus\footnote{Here, we use Higdon's et al. (\cite{higdon}) suggestion about
the galaxy disk orientation, namely that its western half is closest to the
observer.} with velocities of $50-70 \km$, without assuming any circular
rotation. Such outflows are typical of Seyfert galaxies, nevertheless, no AGN
or starburst-like features are observed in the optical spectrum of the galaxy
nucleus (see discussion in Higdon et al., \cite{higdon}). A more reasonable
assumption is that the gas, at $r < 6-7''$, rotates in a polar plane with
respect to the global galactic disk. In this case, the polar gaseous disk lies
roughly in the smallest principal-plane cross-section of the outer bar,
orthogonal to the major axis of the bar, as in \object{NGC 3368}
(Sect.~\ref{sec_3368}).  The hypothesis of a polar disk is supported by the
fact that \object{NGC 5850} has undergone a recent collision with the nearby galaxy
\object{NGC 5846} (Higdon et al. 1998). Through their interaction, part of the gas
could be transported to polar orbits. What is the inclination of this disk to
the line-of-sight? It is easy to show that the angle between the inner disk
plane and the outer galactic disk, $\Delta i$, satisfies the following
relation:

\begin{equation}
\cos \Delta i =\pm\sin i_1 \sin i_0 \cos (PA_1-PA_0) +\cos i_1 \cos i_0,
\end{equation}

\noindent where $i_1$ is the angle between the line-of-sight and the plane
perpendicular to the disk, and $PA_1$ is the position angle of the major axis
of the disk. By substituting $PA_0=335^\circ$, $i_0=37^\circ$
(Table~\ref{tab_pai}), and $PA_1=PA_{dyn}=35^\circ$ into this expression, we
predict  that the disk will be polar ($\Delta i =90\pm5^\circ$) for
$i_1=65^\circ-74^\circ$. The gas rotation curve in Fig.~\ref{5850m} was
calculated for $i_1=65^\circ$.

 At $r>5''$, the gas rotation velocities may be lower than the stellar
ones, but both velocities are consistent within the errors. An extrapolation
of the HI rotation curve (Higdon et al., \cite{higdon}) to the inner radii
gives $V_{rot}\approx30-45\km$ at $r=5''-8''$, in good agreement with our
ionized gas measurements.

\subsection{\object{NGC 5905} }

\label{sec_5905}

Unfortunately, we have not obtained the image of the outer isophotes, extended
beyond a large-scale bar at $r\approx37''$ (the bar size is taken from Friedli
et al., \cite{fri96}). Therefore, we accepted  $i_0=40^\circ$, which was
derived by van Moorsel (\cite{van}) from the HI kinematics analysis. Also we
calculated $PA_0=(130\pm5)^\circ$ from inspection of the HI isovelocities map
(Fig.~8 in van Moorsel, \cite{van}). The last value agrees perfectly with the
results of deep CCD-photometry by Kent (\cite{kent}).

The \Hb velocity fields of stars and ionized gas exhibit  \pak deviating from
$PA_0$ at $r=1''-9''$, in the opposite direction with respect to the $PA$s of
the inner isophotes. Hence they should be caused by non-circular motions in the
large-scale bar. The observed slower [OIII] ionized gas rotation compared with
the \Hb value and the difference of the \pak values in these lines are most
likely connected with shock fronts in the bar edges, as in the case of
\object{NGC 3786} (Sect.~\ref{sec_3786}).

Friedli et al. (\cite{fri96}) showed that the turn of the inner isophotes, at
$r<10''$, is caused by a secondary bar. Nevertheless, they used an obviously
wrong value of $PA_0=45^\circ$. Note that a relatively small value of the
ellipticity of the inner isophotes, and the coincidence of the isophotal $PA$
with $PA_0$ does not allow us to explain the central structure with a
secondary bar. Most probably, the twist of the central isophotes at
$r=5''-10''$ is caused by the transition of the surface brightness
distribution character from a bar-dominated  to a central prolate bulge or
circumnuclear disk. This situation is similar to that of \object{NGC 5566}
(Sect.~\ref{sec_5566}), but without introducing the mini-spirals.

\subsection{\object{NGC 6951}}

\label{sec_6951}

The orientation parameters from Table~\ref{tab_pai} agree with
P\'erez's et al. (\cite{perez}) results. The ionized gas
velocities, measured by them in three different positions of the
slit, coincide completely with our velocity fields.
However, we did not detect the two kinematic components in stellar
rotation, perhaps because of the low MPFS spectral resolution,
relative to that of P\'erez et al. (\cite{perez}) long-slit data.
In the star velocity field, the \pak differs, by $5^\circ-10^\circ$,
from the $PA_0$ (Fig.~\ref{6951m}). The turn of the dynamical axis
is not very significant. This fact is connected with non-circular
motions in the external bar, because the bar isophotes deviate
in the opposite direction relative to $PA_0$.

At large distances from the center, the IFP \Ha and [NII] velocity fields are
similar (Fig.~\ref{6951ifp}). Nevertheless, the [NII] rotation velocities are
lower than those of the \Ha  inside the circumnuclear starforming ring at
$r=5''-10''$, while the IFP \Ha and [NII] and the MPFS \Hb \pak variations are
the same. In contrast, the MPFS [OIII] velocity field exhibits more
significant non-circular ionized gas motions. The dynamical axis, found from
this line measurements, turns by $\sim30^\circ$ at $r<3''$, and the rotation
velocity is lower than the \Hb velocities by a factor of 2. The intensity of
the [OIII] emission decreases sharply outside this region. It is possible that
radial motions of highly excited ionized gas, with a speed of the same order of
the rotation velocities, occur here. Radial motions are connected with the
outflow from an active nucleus, which is of  a transition type between a LINER
and a Sy~2 (P\'erez et al., \cite{perez}). Moreover, the non-thermal radio
continuum image shows a jet-like feature, extended from the nucleus up to
80~pc (Saikia et al., \cite{saikia}). Unfortunately, the MPFS data cannot
resolve this asymmetric  structure.

There is a circumnuclear ring  at $r=5-7''$, nested in the large-scale bar,
which has a $r\approx60''$. The turn of isophotes, in the ground-based optical
images, inside $r<6''$ was explained by Wozniak et al. (\cite{woz}) with a
possible secondary bar. However, Friedli et al. (\cite{fri96}), based on
NIR-photometry, have assumed that the isophote  distortion is caused by a
complex distribution of dust and star-forming regions inside the central
kiloparsec. The HST image resolved this ring into individual starforming
knots. The ring seems to be elliptical, but it is almost circular after
deprojection onto the galactic plane ($PA$ and $\epsilon$ of the central
isophotes correspond to the outer disk orientation). Indeed, the HST
high-resolution images confirm the lack of a secondary bar as an elongated
stellar structure (Fig.~\ref{6951m}).

The HST residual brightness maps reveal a multi-armed flocculent spiral
(Fig.~\ref{6951m}), described in detail  by P\'erez et al. (\cite{perez}). The
mini-spiral is probably associated only with the distribution of gas and dust
rather than with the stellar component. As   has been shown by P\'erez et al.
(\cite{perez}), the nuclear spiral structure, clearly seen in the V band,
completely disappears in the H band, where the effect of dust absorption is
much weaker.

Figs.~\ref{6951m} and \ref{6951ifp} reveal that the position angle of the
dynamical axis, calculated from the \Ha, \Hb, and [NII] velocity fields, at $r
= 0-8''$ deviates by $10-15^\circ$ from the $PA_0$. This    proves the
significant role of non-circular motions in the disk kinematics. Since there
is no inner bar in this region, we can offer the following interpretation of
the observed picture. A gaseous dusty disk of $6''-8''$ ($400-600$ pc) in
radius is embedded into the large-scale bar. In this disk, a multiarmed spiral
structure has developed and perturbs the circumnuclear gas rotation. The
dynamical decoupling of this disk is also confirmed by a high central
molecular-gas density (Kohno et al., \cite{koh}), and by the location of two
ILRs of the large-scale bar (P\'erez et al., \cite{perez}).

Based on their IFP observations, Rozas et al. (\cite{rozas}) presented similar
arguments to explain the kinematic decoupling of the inner ionized-gas disk in
\object{NGC 6951}. However, they tried to interpret the gas non-circular motions in
terms of an inclined disk or a dissipation of the secondary bar. In a
forthcoming paper we are going to consider a detailed kinematics and
circumnuclear structure of this galaxy, based on new 2D-spectroscopy data,
with  sub-arcsecond angular resolution (Moiseev, private communication).

\subsection{\object{NGC 7743}}

\label{sec_7743}

A regular two-armed spiral is seen in the ground-based images, most clearly in
residual images (Fig.~\ref{7743m}). The turn of the isophotes, inside
$r\approx50''$, was explained by Wozniak et al. (\cite{woz}) as the combined
influence of a bar and a triaxial bulge, with the bulge dominating at
$r<10''$. Our analysis of residual images  and the behavior of the $m=2$
Fourier harmonic show  that the bar has a smaller size, and that the spiral
structure distorts the elliptical isophotes at $r>20''-30''$. The sharp
isophote twist, at $r<3''$, can be seen as well in ground-based as in HST
images (see Fig.~\ref{7743m}). The WFPC2 residual optical images reveal a
complex multi-armed spiral (Fig.~\ref{7743m}).

This spiral completely disappears in the near-infrared NICMOS images.
Apparently, this spiral has a dust origin (like the gaseous-dusty spiral in
\object{NGC 6951}). Unfortunately, our MPFS ionized-gas velocity field, in the
mini-spiral region, is uncertain. The [OIII] emission line is weak and noisy.
We can only conclude that at $r<2''-4''$ the gas follows the stellar rotation.
A region with non-circular motions is located at large distances, South from
the center .

The velocity field of the stars shows a regular rotation with almost constant
\pak, while the mean \pak value differs, by $\sim30^\circ$, from $PA_0$. The
deviation  goes in the same direction as the turn of the inner isophotes.
Moreover, the \pak$=300-310^\circ$ coincides with the orientation of the inner
NICMOS isophotes. This   is normal for an inclined rotating nuclear disk.
Nevertheless, on the our ground-based and HST images  this region ($r<8''$) is
not distinguish, except by the mini-spiral at $r<4''$. It is possible that the
accepted value of $PA_0$ is not correct. Additional observations of this
galaxy focused on the large-scale gas and stellar kinematics are needed.

The stellar velocity dispersion field is noisy (Fig.~\ref{7743m}), but it is
in good agreement with the long-slit observations of Kormendy
(\cite{kormendy7743}), concerning the values of the central velocity
dispersion, and the position of the $\sigma_*$ peak, shifted by $3''-4''$ from
the photometric center.

\section{Discussion}

 \label{disc}
\subsection{Circumnuclear morphology of the candidate double-barred galaxies}

In  Table~\ref{tab_res} we collect all structural and kinematic features which
have been found in the observed galaxies. The table contains galactic name,
type (according to the RC3 catalogue), accepted distance from the HyperLeda
database ($H_0=75\km\,\mbox{Mpc}^{-1}$), bar major-axis length and position
angle\footnote{ The  values of $a$ and $PA$ were estimated  from our isophotal
analysis using  same method   as in Wozniak et al.(\cite{woz}). Erwin
(\cite{erwcat}) has provided new definitions for bar semi-major axis
measurement.} and brief description of the structures at different radial
scales. The properties for a single bar are given in each case because we
could not confirm kinematically secondary bars in any galaxy. Usually the bars
described were denoted as ``primary'' in the references. However in \object{NGC 3368}
and \object{NGC 3786} only nuclear dynamically decoupled bars are present (denoted as
``secondary'' in the references), and ``large-scale primary bars'' are, in
fact, a consequence of isophotes distorted by spiral arms
(Sect.~\ref{sec_3368} and \ref{sec_3368}).

We have detected  various types of deviations of the stellar and
gaseous  motions  from the normal  regular rotation in  the
circumnuclear regions of the galaxies observed. These types are separated
below.

\begin{table*}
\caption{Circumnuclear structure of the sample galaxies.}
\begin{center}
\label{tab_res}
\begin{tabular}{rcr@{\hspace{12pt}}rr@{\hspace{12pt}}ccl}
\hline\hline
Name & Type & D$^1$   & \multicolumn{2}{c}{Bar} & \multicolumn{3}{l}{Circumnuclear structure}\\
     & (RC3)& Mpc & a,$''$& PA,$^\circ$ & r, $''$ & r, kpc &  type of the feature \\
\hline
\object{NGC 470} & SAb   &32.6& 30 & 20    &   $0-6$ & $0-1.1$      & lopsided nuclear disk \\
        &       &    &    &       &  $8-20$ & $1.3-3.2$    & gas radial motions in the bar \\
&&&&&&&\\
\object{NGC 2273}& SBa   &25.7& 25 & 108   &   $0-3$ & $0-0.4$      & jet outflow in [OIII] line \\
        &       &    &    &       &   $1-4$ & $0.1-0.5$    & nuclear disk with mini-spirals \\
        &       &    &    &       &  $3-27$ & $0.4-3.4$    & gas radial motions in the bar \\
&&&&&&&\\
\object{NGC 2681}&SAB0/a &9.6& 60 & 30    &  $0-5$  & $0-0.2$      & stellar/gaseous polar (inclined) disk\\
        &       &    &    &       &  $5-20$ & $0.2-0.9$    & decoupled disk with dusty mini-spiral\\
&&&&&&&\\
\object{NGC 2950}& SB0   &18.9& 40 & 155   &  --      & --            & -- \\
&&&&&&&\\
\object{NGC 3368}& SABab &10.6& 5  & 125   &   $0-2$ & 0.1          & gaseous polar disk \\
        &       &    &    &       &   $0-5$ & $0-0.3$      & stellar radial motions in the mini-bar \\
        &       &    &    &       & $5-200$ & $0.3-11$     & global gaseous  disk inclined to the stellar one \\
&&&&&&&\\
\object{NGC 3786}& SABa  &36.1& 6  &  70  & $2-6$   &  $0.4-1.0$   & gas/stellar radial motions in the mini-bar \\
&&&&&&&\\
\object{NGC 3945}& SB0   &17.7& 40 &  72   &  $0-6$  & $0-0.5$      & counter-rotating gaseous disk \\
&&&&&&&\\
\object{NGC 4736}& SAab  & 4.8& 18 & 20    &  $0-25$ & $0-0.6 $      & mini-spiral structure with non-circular gas motions\\
&&&&&&&\\
\object{NGC 5566}& SBab  &19.9& 20 & 155   &  $0-8$  & $0-0.8$      & nuclear disk with mini-spiral \\
&&&&&&&\\
\object{NGC 5850}& SBb   &34.2& 80 & 115   & $0-6$   & $0-1.0$      & polar gaseous disk\\
&&&&&&&\\
\object{NGC 5905}& SBb   &47.2&$ 30$& 200  & $0-8$   & $0-1.9$      & nuclear disk; gas/stellar  radial motions in the bar\\
&&&&&&&\\
\object{NGC 6951}& SABbc &21.9& 80 & 90    & $0-3$   &  $0-0.3$      & jet outflow in [OIII] line\\
        &       &    &    &       & $1-8$   &  $0.1-0.8$    & stellar  radial  motions in the bar\\
        &       &    &    &       & $3-8$   &  $0.3-0.8$    & nuclear disk with mini-spiral and non-circular gas motions\\
        &       &    &    &       & $30-55$ & $3.2-5.9$     & gas radial motions in the bar \\
&&&&&&&\\
\object{NGC 7743}& SB0   & 24.4& 30? & 95  & $0-4$  & $0-0.5$       & mini-spiral structure \\
        &      &    &     &       & $0-8$  & $0-8$       & warped (inclined) disk?
        Need more information  \\
\hline
\end{tabular}
\end{center}
\end{table*}

\subsection{Non-circular motions in bars}

The \pak calculated over the ionized gas velocity fields disagrees with the
$PA_0$ of the lines of nodes in all galaxies where emission lines have been
detected. In Sect.~\ref{atlas0} we have shown that the central turns of
ionized-gas \pak occur in opposition  to the ``primary bar'' isophotes
independent of the orientation of the secondary bar-like structure (excluding
galaxies with inner polar and counter-rotated gaseous disks, Sect.~\ref{polar}
and \ref{counter}). We  explain such behaviour  of the ``dynamical axes''  as
non-circular gas motions (radial flows) in a large-scale bar (or   nuclear
mini-bars in the case of \object{NGC 3368} and \object{NGC 3786}) in agreement with models
described  in Sect.~\ref{method}. In the stellar velocity fields the
amplitudes of \pak deviations are usually less than those of the gas; in
half of the sample \pak is close to the $PA_0$ value. In \object{NGC 3368}, \object{NGC 3786},
\object{NGC 5905}, and \object{NGC 6951} significant (by $7-20^\circ$) \pak deviations are
detected.  Again comparing these with the isophotal $PA$ they are typical signatures  of
non-circular motions of the stars in a single-barred galaxy, or of radial
motions due to elongated stellar orbits, including the case of twisted $x_2$
orbits.

 \object{NGC 2950} is an interesting case, where  the inner isophotes
are turned by more than $50^\circ$ (Fig.~\ref{2950m}), but the \pak
for the stellar component is aligned with the line of nodes.
Neither the outer bar nor the inner bar-like structure have an
appreciable effect on the stellar velocity field. It is  possible
that both disk stellar motions within the bar and the rotation of
the bulge, whose contribution in this SB0 galaxy must be
significant, are observed along the line-of-sight. Since the
spectral resolution was too low to separate two dynamical
components within the line-of-sight velocity distribution (LOSVD),
the mean velocity field corresponds to circular rotation. The
stellar component associated with the outer bar in \object{NGC 2950}
affects the velocity dispersion distribution, whose increase
(elliptical peak of $\sigma_*$) implies a broadening of the LOSVD
in the bar region (Sect.~\ref{dispdis}).

\subsection{Embedded disks?}
\label{disks}

In a significant fraction of the galaxies   inner disks nested in the
central regions of the large-scale bars are found from kinematical and
morphological analysis (in \object{NGC 470}, \object{NGC 2273}, \object{NGC 2681}, \object{NGC 5566},
\object{NGC 5905}, and \object{NGC 6951})\footnote{Here we do not mention \object{NGC 3945} which has
a massive inner disk embedded in the bar, discussed by Erwin et al.
(\cite{erwet03}). In his last paper based on photometric data Erwin
(\cite{erwcat}) also argued  an inner disk origin for the isophotal twists
in \object{NGC 470} and \object{NGC 2273}.}. Below we summarize the signatures indicating
this inner-disk structure:

\begin{itemize}

\item \textit{Photometry}: an orientation of the inner isophotes agrees
with   the $PA_0$ of the line of nodes of an outer disk.

\item \textit{Kinematics:} the central \pak of the stellar velocity field lies
near    the $PA_0$.

\item \textit{Morphology:} There are inner mini-spirals at distances corresponding
to the nuclear disk ($r<1$ kpc) in \object{NGC 2273}, \object{NGC 2681} \object{NGC 5566}, and \object{NGC 6951}. The origin
of the mini-spirals is not understood yet; various models are proposed (for
review see  Elmegreen et al., \cite{elmegreen}; Englmaier \&  Shlosman,
\cite{engl}).  Nevertheless, a ``spiral'' is some disk instability developed in
the relatively thin and flat layer. The mini-spirals also appear  in the images
of \object{NGC 4736} and \object{NGC 7743}, though we cannot determine inner disks in these
galaxies.

\item \textit{Resonances:} In the case of \object{NGC 2681} and \object{NGC 6951} we have shown
that an inner disk region is kinematically  decoupled because it lies
inside the ILR of a global bar.

\end{itemize}

To confirm
the inner disk hypothesis we tried to extract their  brightness
distribution in the galactic plane. For this goal we subtracted a
large-scale model of galaxy as a whole from the available images.
A two-dimensional model of the
surface brightness distribution is needed, because barred galaxies are actually
non-axisymmetric. Our model included a disk with central brightness $\mu_d$ and
exponential scale $r_d$, S\'ersic (\cite{sers}) bulge and global bar. A
two-dimensional surface distribution of the bulge brightness is characterized
by three parameters: effective brightness $\mu_e$,  radius $r_e$, and apparent
ellipticity $\epsilon_b$ (see Eq.~(1) in Moriondo et al., \cite{mor}).
For a bar surface brightness distribution we adopted Ferrer's ellipsoidal
distribution (Binney \& Tremaine, \cite{bin}) projected onto the sky plane as:

\begin{equation}
I(x,y)=I_b(1-(x/a)^2-(y/b)^2)^m
\end{equation}

(the bar is aligned with the x-axis). Here $a$ and $b$ are the semi-axes of the bar,
$m$ is an integer parameter, $I_b$ (or $\mu_b$ when given in
magnitudes) is the central brightness. Also the position angle of the bar  $PA_b$ must
be added  for an arbitrary coordinate axes orientation.

The model was calculated through the iterative technique by means of the
IDL-based software GIDRA (Moiseev, \cite{gidra}). As a first step we
average the original image data in the ellipses with $PA_0$ and $i_0$ and
calculate a first approximation of the $m_d$ and $r_d$ in the
disk-dominated region from the brightness profile obtained. The 2D model
of the exponential disk is subtracted from the original image. The
bar-dominated region of the residual images then is interactively fitted
for the determination of the parameters of the bar. After removal of the 2D
model of the bar we accept the ellipticity of the isophotes as a first approximation
for the bulge ellipticity $\epsilon_b$ and construct the averaged profile again. This
profile is fitted by   Se\'rsic's law. Then the cycle is repeated to
refine the model. Usually 2-3 iterations are enough for a realistic
model.  The model-subtracted images and the accepted parameters of the
models are shown in Fig.~\ref{2Dprf}.  We note that the profiles in the
right plots of Fig.~\ref{2Dprf} are calculated by averaging all component
images, these are not a ``classical'' one-dimensional fitting! Also we have
deprojected the residual images to the galactic plane (Fig.~\ref{2Dprf}, middle
column). The nuclear over-subtraction for \object{NGC 470}  relates to the
atmospheric seeing limitation. The deprojected images reveal positive
brightness residuals as   true circular structures around the nucleus
of \object{NGC 470} and   spiral structures in the inner round disks in \object{NGC 2273}
and \object{NGC 5566}. The case of \object{NGC 2681} is more complex, the residual disk is
elliptical in the galactic plane, which can be connected with a turn of
$x_2$ orbits inside the ILRs. Unfortunately there is no 2D kinematic
information for $r=10-20''$, new observational data  are needed for a more definite
conclusion about this galaxy.

The main conclusion is that \textit{the residual brightness distribution
shows a disk-like morphology} (including ring-like one in \object{NGC 470}).
Note that our decomposition has a preliminary character. More
complex 2D models combining   images in different bands (optical,
NIR) and with different resolutions (HST plus ground-based) is
required for a detailed inspection of the inner disks in barred galaxies.
But such modelling  is beyond the scope of our paper.

\begin{figure*}
\centering
 \caption{Surface brightness decomposition for four galaxies.
 \textbf{Left column} --  the 2D  model was removed from the
 original  image. The solid line corresponds to the galactic major axis.
 \textbf{Central column} -- deprojection of the residual brightness    to the
 galactic plane. The gray scale (common for the left and central images)
 is in magnitudes. Circles mark the radius of the inner disks denoted in
 Table~\ref{tab_res}.
 \textbf{Right column} -- surface brightness profiles. Solid lines:
 azimuthally averaged  profiles of the images; dotted lines:
the same for the components (bulge, bar, and exponential
 disk) of the model. The parameters of the components are also  given here.}
 \label{2Dprf}
\end{figure*}

\subsection{Polar (inclined) disks}

\label{polar}

The kinematic data provide evidence for inner polar disks in \object{NGC 2681} and
\object{NGC 5850}. Significant deviations (by more than $50^\circ$) of the \pak
from the line of nodes occurring in the same direction as the $PA$ of the
inner isophotes are  used as the main argument for this conclusion. Also a
similar disk in \object{NGC 3368} is detected on the basis of the HST morphology
(the dust lane corresponds to the edge of the disk etc., see
Sect.~\ref{sec_3368} and Sil'chenko et al., \cite{leo}) because the
angular resolution of our 2D spectrographs fails at these spatial scales.
What is the origin of these disks? We see two  possibilities.

Firstly, the disks could be formed by an accretion of external matter with
specific orientation of the angular moment, as in ``classical'' large-scale
stable polar rings (see, for example, Arnaboldi \& Sparke, \cite{polar_t}).
Indeed, \object{NGC 2681} demonstrates some spectral (widely distributed starburst) and
structural (disturbed dust lanes in the central disk) features, which have been
ascribed to the merging of external gas (see Cappellari et al., \cite{cap} and
Sect.~\ref{sec_2681}). It is possible that in \object{NGC 3368} all
ionized/neutral/molecular gas collected in the global warped disk  has an
external origin (Sect.~\ref{sec_3368} and Sil'chenko et al., \cite{leo}), which is
is implied by the large-scale distribution of the neutral gas in the Leo~I
group (Schneider, \cite{hi3368}). Also the morphology of the neutral gas in
\object{NGC 5850} suggests a high-speed encounter with the nearby \object{NGC 5846} (see
Sect.~\ref{sec_5850} and Higdon et al., \cite{higdon}).

For the second point of view, we must note that the polar nuclear disks in
\object{NGC 3368} and \object{NGC 5850} are orthogonal to the major axes of the bars.  In recent
years, such polar mini-disks associated with a large-scale bar or a
triaxial bulge have been detected in the circumnuclear regions of several
isolated galaxies, for example in \object{NGC 2841} (Afanasiev \& Sil'chenko,
\cite{afanasil}) or \object{NGC 4548} (Sil'chenko, \cite{sil_polar}). It is
possible that the gas in the centers of these galaxies has been moved
into   polar orbits due to the effect of the bar as   is suggested in a
qualitative model by Sofue \& Wakamatsu (\cite{cof94}). Unfortunately,
detailed numerical simulations of this process and a check of its stability
are absent.

It is possible that both mechanisms (internal as well as  external ones)
are at work in the described galaxies. For a detailed discussion on polar disks
with  sizes of a few hundred parsecs see Sil'chenko (\cite{sil_polar}) and
Corsini et al. (\cite{corsini}). In the last paper a list of 15 galaxies with
 nuclear polar disks  has been collected  and their properties were
investigated. New observations and numerical simulations are needed to
understand completely these galactic structures. We hope that increasing
amounts of  observational information about inner polar disks
will stimulate new theoretical work  in this field.

\subsection{A counter-rotation}
\label{counter}

We have found that the ionized gas in the central kiloparsec of
\object{NGC 3945} rotates in the opposite direction with respect to the
stellar disk. According to Kuijken et al. (\cite{kui}), the
gaseous disks in lenticular galaxies demonstrate a
counter-rotation phenomenon in   $24\pm 8\%$ of all cases.
Therefore, this is a frequent kinematic feature in early-type
galaxies. This   is probably attributable to a merger of an
accreted gaseous cloud with the corresponding direction of
angular momentum  (Bertola et al., \cite{bertola92}). Against this
background, counter-rotation of the gas in the center of \object{NGC 3945},
one of four S0 galaxies in our sample (see Table~\ref{tab_res})
and one of three   with gaseous disks (\object{NGC 2681}, \object{NGC 3945}, and
\object{NGC 7743}), comes as no surprise.

\subsection{Stellar velocity dispersion appearance}
\label{dispdis}

Simple models of single-barred galaxies have shown (Miller \& Smith,
\cite{mil}; Vauterin \& Dejonghe, \cite{vd}) that the central ellipsoidal
peak in the distribution of $\sigma_*$ must be aligned with the orientation
of the outer bar. In our sample the ellipsoidal peaks of the $\sigma_*$
distributions are found in six galaxies (\object{NGC 470}, \object{NGC 2273}, \object{NGC 2681},
\object{NGC 2950}, \object{NGC 3945}, \object{NGC 5905}). As   has been shown in our previous
papers (Moiseev et at., \cite{mexic}; Moiseev, \cite{mois02b}), over our
sample the position angle of the $\sigma_*$ peaks reveals a strong
correlation with the $PA$ of the major axis of the large-scale bar   (see Fig.~4 in
Moiseev, \cite{mois02b}). But there is no  any correlation between the
symmetry axis of the velocity dispersion and the inner isophotal
orientation. So in these objects the large-scale bars determine the
dynamics of the stellar component even in the regions where the
photometric ``inner bars'' are observed.

In \object{NGC 3786} a central drop of the stellar velocity dispersion is found
(Sect.~\ref{sec_3786}). Recently, Emsellem et al. (\cite{emsellem01}) and
M\'arquez et al. (\cite{drop}) have observed similar drops in the
$\sigma_*$ radial distributions in the central parts of some barred
galaxies with AGN. These drops can probably be attributed to the ``cold''
(from the dynamical point of view) stellar disk embedded into the bar.
Indeed, the recent self-consistent N-body simulations by Wozniak et al.
(\cite{woz03}) show that ``young stars born in the nuclear regions from
dynamically cold gas have a lower velocity dispersion than the older
stellar populations''. Such a drop was also previously observed in \object{NGC 6951}
(P\'erez et al., \cite{perez}; M\'arquez et al., \cite{drop}) up to
distances of $\pm6-8''$ from the center  (Fig.~5 in M\'arquez et al.,
\cite{drop}). Unfortunately, our $\sigma_*$ map for this galaxy has a
smaller size (Fig.~\ref{6951m}), so we could not detect this interesting
feature of the stellar kinematics.

\subsection{Molecular gas morphology}

\label{co}

Recently Petitpas \& Wilson (\cite{petit}, \cite{petit03}) proposed a new
method for resolving the origin of the secondary bar. They compared turns of NIR
isophotes with molecular gas (CO) distributions. Namely, if NIR and CO
images have similar features, this may suggest that the isophote twists are
related to structures hosted by disks and not to triaxial bulges.  In
our sample, we have found detailed CO-maps for \object{NGC 470} (Jogee et al.,
\cite{jog}), \object{NGC 2681} (Jogee et al., \cite{jog}), \object{NGC 2273} (Petitpas \&
Wilson, \cite{petit}), \object{NGC 3368} (Sakamoto et al., \cite{sakamoto}),
\object{NGC 4736} (Wong \& Blitz, \cite{wong}), and \object{NGC 6951} (Kohno \cite{koh}).
In five galaxies (except \object{NGC 3368}) the orientation of the CO-distribution
coincides with the isophotes of our NIR (or optical) images. We think that the
physical reasons for this effect can be different:

\begin{itemize}

\item In \object{NGC 470}, \object{NGC 2273}, and \object{NGC 6951} the iso-contours  of the inner
CO distribution match  the $PA$ and $\epsilon$ of the outer isophotes.
We suggest that molecular gas is distributed in the nuclear disks
discussed above (Sect.~\ref{disks}), as the radial  scale of the CO
concentration  and the photometric disks are the same. Two gaseous
clumps, symmetric with respect to the center found by Petitpas \& Wilson
(\cite{petit}) in \object{NGC 2273}, can be connected with bright segments of the
inner mini-spiral (Fig.~\ref{2Dprf}), or may correspond to the stellar
and gas orbit crowding near the ILR of the primary bar (see Petitpas \&
Wilson, \cite{petit}). Moreover, in the center of \object{NGC 6951} the molecular gas
distributed in the pseudo-ring agrees with the star-forming ring visible in
optical images (Sect.~\ref{sec_6951}). But in contrast with the optical bands,
the CO emission is dominated by a ``twin peaks'' morphology,  which is also
interpreted by Kohno (\cite{koh}) as ``$x_1/x_2$ orbit-crowding regions''.
Therefore, the distribution of the molecular gas  in the  discussed
galaxies can  differ from purely  disk-like, because  these disks nest in the
non-axisymmetric potential of a large-scale bar. Nevertheless, the CO
emission  is concentrated on the same scales as the photometric nuclear disks.

\item In \object{NGC 2681} and \object{NGC 3368} the central molecular gas is confined to
the inner polar disks, hence these disks are CO-rich. But in \object{NGC 3368}
the NIR isophotes cannot be related to the polar disk, because this disk
is seen in the dust distribution only, not in that of the stellar component
(see scheme in Fig.~18 in Sil'chenko et al., \cite{leo}).

\item In \object{NGC 4736} there is a molecular bar present elongated along the stellar
bar (see notes in  Sect.~\ref{sec_4736}).

\end{itemize}

So  the molecular gas morphology does not contradict the existence of the
structures found in this paper (nuclear coplanar and polar disks).

\subsection{Where are the secondary bars?}

The circumnuclear regions (of a few kiloparsec in size) in all candidate
double-barred galaxies of our sample (except \object{NGC 5566}) are dynamically
decoupled, because  deviations of the gaseous/stellar dynamical axes
relative to the lines of nodes or elliptical peaks of $\sigma_*$ are
detected on our 2D maps. Various peculiar structures (inner polar and
coplanar disks, mini-spirals) were revealed. Such a large fraction of
peculiar structures is not surprising, because different structural
features can produce   twists of the inner isophotes (see Introduction).
However, we have not found any kinematic features connected with inner
(secondary) bars. This cannot be explained by the limited angular
resolution of our kinematic data, because we confirm definitively the nuclear
mini-bars (without large bars) in \object{NGC 3368} and \object{NGC 3786}. We see two
possibilities to explain this.

Firstly, it is possible that the twist of the inner isophotes of galactic
images has \textit{no connection with a dynamically decoupled inner bar}.
In this case a ``secondary  bar'' is an illusive structure, and
different reasons may cause the isophotal distortion: inner disks or
rings, nuclear spirals etc. Indeed, Shaw et al. (\cite{shaw}) had
constructed a model where the inner twist of the stellar isophotes is
triggered by a secular evolution of the gaseous component  in a
single-barred  gas-rich galaxy. See also papers by  Friedli et al.
(\cite{fri96}), Moiseev (\cite{mois01b}), and  Petitpas \& Wilson,
(\cite{petit}) where various mechanisms are considered for an isophotal
twist explanation.

Secondly, the kinematics of the inner bar in a double-barred galaxy
\textit{differs essentially  from  single-bar dynamics}. In this case, the
deviations of \pak and the peaks of the velocity dispersion  must be interpreted in a
different manner than in this paper. For example, in their last work Corsini
et al. (\cite{cor2950}) have measured different pattern speeds for secondary
and primary bar-like structures in \object{NGC 2950}. However, \pak of the stellar
velocity field coincides with the line of nodes (see Sect.~\ref{sec_2950}).
New self-consistent models of barred galaxies must be used to investigate this
phenomenon.

Of course, selection effects must be kept in mind because our sample is
small: we studied only 13 galaxies from Moiseev's (\cite{mois01b}) list,
or only 6 objects from 50 ``photometrically  confirmed'' double-barred
galaxies of Erwin (\cite{erwcat}). However, our sample contains the most
``classical'' double-barred objects with  strong twist of the inner
isophotes (\object{NGC 2681}, \object{NGC 5850}).

Nevertheless, the existing samples of double-barred candidates must be revised.
First of all, additional 2D kinematics data (panoramic spectroscopy, HI and CO
interferometric radio observations) must be acquired for studying any
dynamical decoupling of the circumnuclear regions. We hope that our paper is
only the first small cautious step in this promising way.

\section{Conclusion}

\label{concl}

The  analysis of the morphology and kinematics of 13 double-barred galaxy
candidates observed by means of  panoramic spectroscopy at the SAO RAS
6m telescope is presented in this paper. We tried to collect the maximum
possible information from observed 2D distributions of the optical and NIR
surface brightness and kinematic properties. For this goal, panoramic
spectroscopy observations were carried out at the 6m telescope. Velocity fields and velocity dispersion
fields of ionized gas and stars  were
constructed from these observations. The rotation curves and radial
variations of the dynamical axis position angle were calculated. The
analysis of the morphology of the galaxies was done by means of isophote
analysis of NIR and optical ground-based and HST archival images.

We suggest that candidate double-barred galaxies are, in fact, galaxies with
\textit{very different circumnuclear structure}. It is necessary to note that
the majority of the observed  morphological and kinematic features in our
sample galaxies may be explained without the secondary bar hypothesis. Three
cases of inner polar disks, one counter-rotating  gaseous disk and seven nuclear
disks (with and  without mini-spirals) nested in   large-scale bars
were found in this work. However, we think that the observational data
presented here, particularly the 2D data, will be useful to proponents
as well as to opponents of the secondary bar hypothesis. Doubtless  the
models of such galaxies must explain all their observed kinematic properties.

\begin{acknowledgements}
We would like to thank Olga~Sil'chenko, Victor Afanasiev, Lilia
Shaliapina and Enrico Corsini for their interest and numerous
discussions. We also thank the referee, Glen Petitpas for careful reading
and useful comments. This research has made use of the NASA/IPAC
Extragalactic Database (NED), which is operated by the Jet Propulsion
Laboratory, California Institute of Technology, under contract with the
National Aeronautics and Space Administration. Also HyperLeda database
and data from Canadian Astronomy Data Center (CADC) were used. CADC is
operated by the Herzberg Institute of Astrophysics, National Research
Council of Canada. The authors would like to thank all the staff of the
{\it Observatorio Astron\'omico Nacional}, in San Pedro Martir, M\'exico
for their assistance during the NIR observations. This work was partially
supported by the RFBR grants No.~03-02-06070mas, 02-02-16878 and by the
Russian Science Support Foundation.
\end{acknowledgements}

{}


\begin{thebibliography}{}
\bibitem[2003]{fast} Aguerri, J.A.L.,  Debattista, V.P.,  Corsini,
E.M., 2003, \mnras, 338, 465
\bibitem[2001]{mpfs} Afanasiev, V.L., Dodonov, S.N., Moiseev, A.V., 2001, in
``Stellar dynamics: from classic to modern'', eds. Osipkov L.P., Nikiforov
I.I., Saint Petersburg, 103
\bibitem[1998]{trans} Afanasiev, V.L., Mikhailov, V.P., Shapovalova, A.I.,
1998, Astron. Astrophys. Trans, 16, 257
\bibitem[1981]{af_shap81} Afanasiev, V.L. \& Shapovalova, A.I., 1981, Astrophysics, 17, 221
\bibitem[1996]{af_shap96} Afanasiev, V.L. \& Shapovalova, A.I., 1996,  ASP Conf. Ser., 91, 221
\bibitem[1999]{afanasil} Afanasiev, V.L., \& Sil'chenko, O.K., 1999, \aj, 117, 1725
\bibitem[1994]{polar_t} Arnaboldi, M. \& Sparke, L.S., 1994, \aj, 107, 958
\bibitem[2003]{barsel} Barnes, E.I., \& Sellwood, J.A., 2003, \aj, 125, 1164
\bibitem[1989]{begeman} Begeman, K.G., 1989, A\&A, 223, 47
\bibitem[1987]{bender} Bender, R. \& M\"{o}ellenhoff, C., 1987, \aap, 177, 71
\bibitem[1992]{bertola92} Bertola, F., Buson, L.M., Zeilinger, W.W., 1992, \apj, 401, L79
\bibitem[1995]{bertola95} Bertola, F., Cinzano, P., Corsini, E. M., Rix, H.-W., Zeilinger, W.W., 1995, \apj, 448, L13
\bibitem[1987]{bin} Binney, J., \& Tremaine, S., 1987, Galactic Dynamics: Princeton
\bibitem[1996]{buta96} Buta, R., 1986, \apjs, 61, 609
\bibitem[1993]{butacr} Buta, R., \& Crocker, D.A., 1993, \aj, 105, 1344
\bibitem[2001]{cap} Cappellari, M., Bertola, F., Burstein, D., et al., 2001, \apj, 551, 197
\bibitem[1978]{chev} Chevalier, R.A., \& Furenlid, I., 1978, \aj, 225, 67
\bibitem[1995]{comb} Combes, F., \& Gerin, M., 1995, \aap, 150, 327
\bibitem[2003a]{corsini} Corsini, E.M., Pizzella, A., Coccato, L., Bertola, F., 2003a, \aap, 408, 873
\bibitem[2003b]{cor2950} Corsini, E.M., Debattista, V.P., Aguerri, J.A.L., 2003b, \apj, 599, L29
\bibitem[1994]{camila} Cruz-Gonzalez, I., Carrasco, L., Ruiz, E., et al., 1994, Rev. Mex. Astron. Astrofis., 29, 197
\bibitem[1975]{vauc} de Vaucouleurs, G., 1975, \apjs, 29, 193
\bibitem[1995]{dodo} Dodonov, S.N., Vlasyuk, V.V., Drabek, S.V., 1995, ``Interferometer Fabry-Perot: user manual'', N.Arkhiz
\bibitem[2000]{ferruit} Ferruit, P., Wilson, A. S., Mulchaey, J., 2000,
\apjs, 128, 139
\bibitem[2000]{engl}  Englmaier, P., \&  Shlosman, I., 2000, \apj,   528,  677
\bibitem[2004]{erwcat} Erwin, P., 2004, \aap, accepted
\bibitem[1999]{erw99} Erwin, P., \& Sparke, L.S., 1999, \apj, 521, L37
\bibitem[2002]{erw02} Erwin, P., \& Sparke, L.S., 2002, \aj, 124, 65
\bibitem[2003]{erw03} Erwin, P., \& Sparke, L.S., 2003,  \apjs, 146, 299
\bibitem[2001]{erw01} Erwin, P., Vega Beltra\'n, J.C., Beckman, J., 2001, ASP Conf. Proc., 249. 171 (astro-ph/0112056)
\bibitem[2003]{erwet03} Erwin, P., Vega Beltra\'n J.C., Graham, A.W., Beckman, J.E., 2003, \apj, 597, 929
\bibitem[2002]{elmegreen} Elmegreen, D.M., Elmegreen, B.G., Eberwein, K.S, 2002, \apj, 564, 234
\bibitem[2002]{eric470} Emsellem, E., 2002, astro-ph/0202522
\bibitem[2001a]{emsellem01} Emsellem, E., Greusard, D., Combes, F., et al., 2001a, \aap, 368, 52
\bibitem[2001b]{emsellem01b} Emsellem, E., Greusard, D., Friedli, D., Combes, F., 2001b, Astrop. and Space Science (Suppl.), 277, 455
\bibitem[2000]{eric00} Emsellem, E., \& Friedli, D., 2000, ASP Conf. Ser., 197, 51, (astro-ph/9910260)
\bibitem[2004]{monument} Fridman, A.M., Afanasiev, V.L., Dodonov, S.N. et al.,
2004, submitted to \aap
\bibitem[2000]{fridman} Fridman, A.M., \& Khoruzhii, O.V., 2000, Physics Letters A, 276, 199
\bibitem[1996]{frei} Frei, Z., Guhathakurta, P., Gunn, J.E., Tyson, J.A.,
1996, \aj, 111, 174
\bibitem[1993]{frimar} Friedli, D., \& Martinet, L., 1993, \aap, 277, 27
\bibitem[1996]{fri96} Friedli, D., Wozniak, H., Rieke, M., Martinet, L.,
Bratschi, P., 1996, \aaps, 118, 461
\bibitem[1997]{hagen} Hagen-Thorn, V.A. \& Reshetnikov, V.P., 1997, \aap,
319, 430
\bibitem[2001]{heller01} Heller, C., Shlosman, I., Englmaier, P., 2001, \apj,
553, 661
\bibitem[1987]{helou} Helou, G., Salpeter, E.E., Terzian, Y., 1987, \aj, 87,
1443
\bibitem[2001]{hera} Hera\'udeau, P., Simien, F., Maubon, G., Prugniel, P.,
2001, \aaps, 136, 509
\bibitem[1998]{higdon} Higdon, J., Buta, R., Purcel, G.B., 1998, \aj, 115, 80
\bibitem[1991]{gar} Garcia-Gomez, C., \& Athanassoula, E., 1991, \aaps, 89, 159
\bibitem[2001]{jog} Jogee, S., Baker, A.J., Sakamoto, K., Scoville, N.Z., Kenney,
J.D.P., ASP Conf. Proc., 2001, 249, 612 (astro-ph/0201209)
\bibitem[1997]{jung} Jungwiert, B., Combes, F., Axon, D.J., 1997, \aaps, 125,
479
\bibitem[1996]{lopside} Junquera, S., \& Combes, F., 1996, \aap, 312, 703
\bibitem[1984]{kent} Kent, S.M., 1984, \apjs, 56, 105
\bibitem[1999]{koh} Kohno, K., Kawabe, R., Vila-Vilaro, B., 1999, \apj, 511,
157
\bibitem[1982]{kormendy7743} Kormendy, J., 1982, \apj, 257, 75
\bibitem[1996]{kui} Kuijken, K., Fisher, D., Merrifield, M.R., 1996, \mnras,
283, 543
\bibitem[2002]{laine02} Laine, S., Shlosman, I., Knapen, J.H., Peletier, R.F.,
2002, \apj, 567, 97
\bibitem[1999]{lind} Lindblad, P.O., \aap Rev., 1999, 9, 221
\bibitem[1997]{lyakh} Lyakhovich, V.V., Fridman, A.M., Khoruzhii, O.V.,
Pavlov, A.I., 1997, Astron. Reports, 41, 447
\bibitem[2000]{mac00} Maciejewski, W., \& Sparke, L.S., 2000, \mnras, 313, 745
\bibitem[2002]{mac02} Maciejewski, W., Teuben, J., Sparke, L.S., Stone, J.M.,
2002, \mnras, 329 502
\bibitem[2003]{drop} M\'arquez, I.,  Masegosa, J., Durret, F., et
al., 2003, \aap, 409, 459
\bibitem[1979]{mil} Miller, R.H., \& Smith, B.F., 1979, \apj, 227, 785
\bibitem[1998]{gidra} Moiseev, A.V., 1998, preprint of the SAO RAS,  134, 1
\bibitem[2000]{mois00} Moiseev, A.V., 2000, \aap, 363, 843
\bibitem[2001a]{mois01a} Moiseev, A.V., 2001a, \bsao, 51, 11
(astro-ph/0111219)
\bibitem[2001b]{mois01b} Moiseev, A.V., 2001b, \bsao, 51, 140
(astro-ph/0111220)
\bibitem[2002a]{mois02} Moiseev, A.V., 2002a, \bsao, 54, 74, (astro-ph/0211104)
\bibitem[2002b]{mois02b} Moiseev, A.V., 2002b, Astronomy Letters, 28, (astro-ph/0211105)
\bibitem[2000]{mm} Moiseev, A.V. \& Mustsevoi, V.V, 2000, Astro. Letter., 26,
665 (astro-ph/0011225)
\bibitem[2002]{mexic} Moiseev, A.V., Valdes, J.R., Chavushan, V.O., 2002, ASP Conf. Ser., 282, 261 (astro-ph/0202192)
\bibitem[1992]{monnet} Monnet, G., Bacon, R., Emsellem, E., 1992, \aap, 253, 366
\bibitem[1998]{mor} Moriondo, G., Givanardi, C., Hunt, L.K., 1998, \aaps, 130,
81
\bibitem[1997]{mulch} Mulchaey, J.S., Regan, M.W., Kundu, A., 1997, \apjs, 110, 299
\bibitem[1995]{mulder} Mulder, P.S., 1995, \aap, 303, 57
\bibitem[2002]{petit} Petitpas, G.R., \& Wilson, C.D., 2002, \apj, 575, 814
\bibitem[2003]{petit03} Petitpas, G.R., \& Wilson, C.D., 2003, \apj, 587, 649
\bibitem[2000]{perez} P\'erez, E., M\'arquez, I., Durret, F., et al., 2000, \aap, 353, 893
\bibitem[1990]{pfenninger} Pfenninger, D., \& Norman, C.A., 1990, \apj, 363, 391
\bibitem[1994]{lops_obs} Richter, O.-G., \& Sancisi, R., 1994, \aap, 290, L9
\bibitem[2002]{rozas} Rozas, M., Relano, M., Zurita, A., Beckman, J.E., 2002, \aap, 386, 42
\bibitem[2002]{saikia} Saikia, D.J., Phookun, B., Pedlar, A., Kohno, K., 2002, \aap, 383, 98
\bibitem[1999]{sakamoto} Sakamoto, K., Okumura, S.K., Ishizuki, S.,  Scoville, N.Z., 1999, \apjs, 124, 403
\bibitem[2002]{eva} Schinnerer, E., Maciejewski, W., Scoville, N., Moustakas, L.A., 2002, \apj, 575, 826
\bibitem[1989]{hi3368} Schneider, S.E., 1989, \apj, 343, 94
\bibitem[1968]{sers} S\'ersic, J.-L. 1968, Atlas de Galaxias Australes (Cordoba: Obs. Astron.)
\bibitem[2002]{u5600}Shalyapina, L.V., Moiseev, A.V., Yakovleva, V.A., Astronomy Letters, 28, 443, (astro-ph/0206397)
\bibitem[1993]{shaw} Shaw, M.A., Combes, F., Axon, D.J., Wright, G.S., 1993, \aap, 273, 31
\bibitem[1989]{shlosman89} Shlosman, I., Frank, J., Begeman, M.C., 1989, Nature, 338, 45
\bibitem[2002]{shlosman02} Shlosman, I., \& Heller, C., 2002, \apj, 565, 921
\bibitem[2002]{sil_polar} Sil'chenko, O.K., 2002, Astro. Letter., 28, 207
\bibitem[2003]{leo} Sil'chenko, O.K., Moiseev, A.V., Afanasiev, V.L., Chavushyan, V.H., Vald\'es, J.R., 2003, \apj, 591, 185
\bibitem[1993]{co93} Sofue, Y., Wakamatsu, K.I., Taniguchi, Y., Nakai, N., 1993, PASJ, 45, 43
\bibitem[1994]{cof94} Sofue, Y., \& Wakamatsu, K.I., 1994, \aj, 107, 1018
\bibitem[1979]{td} Tonry, J., \& Davis M., \aj, 84, 1511
\bibitem[1997]{turn} Turnbull, A.J., Bridges, A.J., Carter, D., 1997, \mnras, 307, 967
\bibitem[1993]{vanbuta} van Driel, W., \& Buta, R.J., 1993, \aap, 245, 1991
\bibitem[1982]{van} van Moorsel, G., 1982, \aap, 107, 66
\bibitem[1997]{vd} Vauterin, P., \& Dejonghe, H., 1997, \mnras, 286, 812
\bibitem[2001]{vega} Vega Beltra\'n, J.C., Pizzella, A., Corsini, E. M., et al., 2001, \aap, 374, 394
\bibitem[2000]{wong} Wong, T., \& Blitz, L., 2000, \apj, 540, 771
\bibitem[1995]{woz} Wozniak, H., Friedli, D., Martinet, L., Martin, P., Bratschi, P., 1995, \aaps, 111, 115
\bibitem[2000]{woz99} Wozniak, H., 2000, ASP Conf. Ser., 197, 75,(astro-ph/9910007)
\bibitem[2003]{woz03}Wozniak, H., Combes, F., Emsellem, E., Friedli, D.,  2003, \aap, 409, 469
\bibitem[2002]{zashop} Zasov, A.V. \& Khoperskov, A.V., 2002, Astronomy Reports, 46, 173
\end{thebibliography}
\end{document}